\documentclass[iop]{emulateapj}
\usepackage{color} 

\usepackage{amsmath}
\usepackage{amssymb}
\usepackage{hyperref}

\newcommand{\eqr}[1]{Eq.\thinspace(#1)}
\newcommand{\pfrac}[2]{\frac{\partial #1}{\partial #2}}

\newcommand{\mvec}[1]{\mathbf{#1}}

\newcommand{\gvec}[1]{\boldsymbol{#1}}

\newcommand{\cbas}[1]{\gvec{\sigma}_{#1}}

\newcommand{\tbasis}[1]{\tilde{\mvec{e}}_{#1}}
\newcommand{\tdbasis}[1]{\tilde{\mvec{e}}^{#1}}

\newcommand{\gke}{\texttt{Gkeyll}}

\DeclareMathOperator{\sech}{sech}
\DeclareMathOperator{\sign}{sign}

\begin{document}

\title[Relativistic Vlasov]{Modeling of Relativistic Plasmas with a Conservative Discontinuous Galerkin Method}

\author{%
James~Juno\thanks{Email address for correspondence: jjuno@pppl.gov}$^{1}$,
Grant~Johnson$^{2}$,
Alexander~Philippov$^{3,4}$,
Ammar~Hakim$^{1}$,
Alexander~Chernoglazov$^{5}$,
Shuzhe~Zeng$^{3,4}$
}
\affiliation{%
$^1$Princeton Plasma Physics Laboratory, Princeton, NJ 08543, USA\\
$^2$Department of Astrophysical Sciences, Princeton University, Princeton, NJ 08544, USA\\
$^3$Department of Physics, University of Maryland, College Park, MD 20742, USA\\
$^4$Institute for Research in Electronics and Applied Physics, University of Maryland, College Park, MD 20742, USA\\
$^5$Institute for Advanced Study, Princeton, NJ 08540, USA\\
}


\begin{abstract}
We present a new method for solving the relativistic Vlasov--Maxwell system of equations, applicable to a wide range of extreme high-energy-density astrophysical and laboratory environments. 
The method directly discretizes the kinetic equation on a high-dimensional phase-space grid using a discontinuous Galerkin finite element approach, yielding a high-order, conservative numerical scheme that is free from the Poisson noise inherent to traditional Monte-Carlo methods. A novel and flexible velocity-space mapping technique enables the efficient treatment of the wide range of energy scales characteristic of relativistic plasmas, including QED pair-production discharges, instabilities in strongly magnetized plasmas surrounding neutron stars, and relativistic magnetic reconnection. Our noise-free approach is capable of providing unique insight into plasma dynamics, enabling detailed analysis of electromagnetic emission and fine-scale phase-space structure.
\end{abstract}


\section{Introduction}
In a diverse array of plasma environments, from the magnetospheres of pulsars \citep{Philippov:2022} to runaway electrons in fusion reactors \citep{Breizman:2019} to ultra-intense laser-plasma experiments \citep{Gonoskov:2022}, we observe plasmas which are not only hot enough and diffuse enough to be collisionless, i.e. described by kinetic theory, but also relativistic. 
These plasmas are some combination of relativistically hot, with temperatures exceeding the rest mass energy of the particular particles, $T_s > m_s c^2$, and moving at relativistic velocities, $v \sim c$ where $v$ is either a bulk velocity of the plasma or a particular particle velocity.
Modeling these plasmas thus requires solving the Vlasov, or Boltzmann, equation for the evolution of the particle distribution function with the necessary modifications to include relativity, 
\begin{align}
    \partial_t f + & \nabla_{\mvec{x}} \cdot \left (\frac{\mvec{u}}{\gamma} f \right ) \notag \\ 
    & + \nabla_{\mvec{u}} \cdot \left (\frac{q_s}{m_s} \left [\mvec{E} + \frac{\mvec{u}}{\gamma} \times \mvec{B} \right ] f \right ) = C[f] + S, \label{eq:VlasovSR}
\end{align}
where $\mvec{u}$ is the spatial component of the four-velocity, $\mvec{u} = \gamma \mvec{v}$, $\gamma = \sqrt{1 + |\mvec{u}|^2/c^2}$ is the particle Lorentz boost factor, $q_s$ and $m_s$ are the particle charge and rest mass respectively, $\mvec{E}$ and $\mvec{B}$ are the electric and magnetic fields respectively, $C[f]$ includes all discrete particle effects such as collisions and radiation reaction, and $S$ is an explicit source term due to, e.g., pair production. 
\eqr{\ref{eq:VlasovSR}}, for any number of plasma species of interest, is then coupled to Maxwell's equations via (four-) velocity moments of the distribution function to obtain the charge and current density used to evolve the electric and magnetic fields and close the system of equations. 

Traditionally, the numerical solution of \eqr{\ref{eq:VlasovSR}} is done with Monte Carlo methods due to the high dimensionality of the Vlasov equation, reducing the six dimensional solve to evolving ``macroparticles'' in (up to) three spatial dimensions, and sampling these particles to construct the distribution function and deposit charges and currents on a grid to couple to a solver for Maxwell's equations---typically referred to as the particle-in-cell, or PIC method. 
This approach has proved historically fruitful, revealing new insights into a diverse array of plasma phenomena, and there now exist a number of production PIC codes for modeling extreme astrophysical and laser-plasma systems \citep{Fonseca:2002,Spitkovsky:2005,Fonseca:2008,Bowers:2009,Cerutti:2013,Germaschewski:2016,Parfrey:2019,Hakobyan:2024,Hakobyan:2025,Galishnikova:2025}. 
However, Monte Carlo methods inevitably suffer from Poisson noise due to the statistics from using a finite number of particles.
Unfortunately, this noise decreases only very slowly, $\propto\sqrt{N_{\rm{ppc}}}$, where $N_{\rm{ppc}}$ is the number of particles per grid cell \citep{birdsallbook}, and thus in some applications where we are interested in precise phase space dynamics and detailed diagnostics of electromagnetic emission, the standard PIC technique can be inadequate for understanding the plasma's evolution. 

The alternative, directly discretizing \eqr{\ref{eq:VlasovSR}} on a phase space grid and numerically integrating the high dimensional partial differential equation for the physical system of interest, would eliminate sources of error from Monte Carlo methods such as noise, but at increased computational cost. 
This increased computational cost has proved a significant deterrent within the field of plasma physics, especially for the relativistic variant of the kinetic equation. 
As computational capabilities have advanced with new hardware and the development of novel algorithms, a number of production codes for both the continuum discretization of the multi-species Vlasov-Maxwell system equations \citep{Vencels:2016, Juno:2018, Roytershteyn:2018, HakimJuno:2020, AllmanRahn:2022, grunwald:2025} and hybrid-kinetic reductions relevant when the electron kinetics can be ignored \citep{Valentini:2007, Kempf:2013, Palmroth:2018} have been developed, but only for low energy density space, astrophysical, and laboratory plasmas. 
While extensions of numerical methods for kinetic equations such as the Lattice Boltzmann method have been applied to relativistic neutral Boltzmann solvers \citep{Ambrus:2022} and reduced geometry relativistic Vlasov-Maxwell solvers for targeted applications \citep{Wettervik:2017, Ye:2025}, extending the bevy of algorithmic developments to relativistic plasmas more generally has proved nontrivial for two principal reasons.

Firstly, significant breakthroughs in robustness were obtained with conservative grid-based methods for the Vlasov equation, but these breakthroughs relied on particular choices of basis functions in finite element \citep{Cheng:2014b, Juno:2018, Kormann:2025} or spectral \citep{Delzanno:2015, Koshkarov:2021, Pagliantini:2023}  methods that allowed one to prove conservation relations such as conservation of energy were implicitly contained in the spatial discretization of the Vlasov equation, and there exists no obvious basis choice when we add relativity where the relevant conservation relations involve integrating non-polynomial functions over velocity space,
\begin{align}
    \mathcal{E}_{NR} = \int \frac{1}{2} m |\mvec{v}|^2 f \thinspace d\mvec{v}, \quad \rightarrow \quad \mathcal{E}_{R} = \int m c^2 \gamma f \thinspace d\mvec{u}.
\end{align}
Secondly, energetic particles are inevitable in simulations of relativistic plasmas, and thus significant phase space extents can be required for practically any simulation of interest. 
Indeed, an advantage of the PIC algorithm is the fundamental adaptiveness of the phase space resolution, with the particle representation of the distribution function naturally filling only regions of phase space with finite phase space density. 
Thus, a grid-based discretization of the relativistic Vlasov-Maxwell system of equations would need algorithmic breakthroughs to obtain a conservative discretization that had flexible velocity-space meshes and could thus model systems of interest without enormous resolution requirements. 

In this paper, we extend a class of modal discontinuous Galerkin (DG) finite element methods to the relativistic Vlasov-Maxwell system of equations. 
We choose modal DG due to the ability to minimize cost while respecting a fundamental requirement of DG kinetic solvers to eliminate aliasing errors in the integration of the weak form to construct a conservative scheme. 
We extend this modal DG approach into a mixed nodal-modal form to include geometric factors for a mapped (four-) velocity space that allows for stretched meshes that extend to sufficiently high energies to capture the dynamics of the most energetic particles in the simulations. 
We demonstrate how this solver is naturally noise-free compared to a Monte Carlo approach in a variety of benchmarks, including pair production discharges and relativistic magnetic reconnection, providing a complementary approach to the traditional PIC algorithm for modeling relativistic, high energy density plasmas.
This grid-based discretization is implemented within the open-source \gke\ simulation framework, and all input files utilized to perform the simulations can be found in \url{https://github.com/ammarhakim/gkyl-paper-inp} and all plots in this paper utilize the \texttt{postgkyl} package \url{https://github.com/ammarhakim/postgkyl} for analyzing \gke\ data.
\section{Results}\label{sec:results}
\subsection{Model Overview}
We begin by defining a mapped velocity grid, 
\begin{align}
    \mvec{u} = \mvec{u}(\eta^1, \eta^2, \eta^3) = u_1(\eta^1) \gvec{\sigma}_1 + u_2(\eta^2) \gvec{\sigma}_2 + u_3(\eta^3) \gvec{\sigma}_3, 
\end{align}
which has tangent and dual vectors, and velocity-space Jacobian,
\begin{align}
    \tbasis{i} & = \partial_{\eta^i} \mvec{u} = \partial_{\eta^i} u_i \gvec{\sigma}_i, \quad \tdbasis{i} = \frac{1}{\partial_{\eta^i} u_i} \gvec{\sigma}_i, \\
    \mathcal{J}_u & = |\partial_{\eta^1} u_1 \thinspace \partial_{\eta^2} u_2 \thinspace \partial_{\eta^3} u_3|,
\end{align}
where the latter follows from the square root of the determinant of the velocity-space metric, which is simply a diagonal tensor for this choice of velocity map. 
Here, $\cbas{i}$ are unit vectors in Cartesian coordinates and thus a Cartesian coordinate system is expressed in this mapping as, 
\begin{align}
    \mvec{u} = u_x(\eta^x) \hat{\mvec{x}} + u_y(\eta^y) \hat{\mvec{y}} + u_z(\eta^z) \hat{\mvec{z}}. 
\end{align}
In these coordinates, and assuming the velocity map is static in time and configuration space, \eqr{\ref{eq:VlasovSR}} in the absence of collisions or sources becomes
\begin{align}
    \partial_t \left (\mathcal{J}_u f \right ) + & \nabla_{\mvec{x}} \cdot \left (\frac{\mvec{u}}{\gamma} \mathcal{J}_u f \right ) \notag \\ 
    & + \nabla_{\gvec{\mathcal{U}}} \cdot \left (\frac{q_s}{m_s} \left [\mvec{E} + \frac{\mvec{u}}{\gamma} \times \mvec{B} \right ] \mathcal{J}_u f \right ) = 0, \label{eq:VlasovSRNonuniform}
\end{align}
where the new divergence operator is defined as
\begin{align}
    \nabla_{\gvec{\mathcal{U}}} \cdot \mvec{a} = \partial_{\eta^i} \left ( \tdbasis{i} \cdot \mvec{a} \right ).
\end{align}
Thus, with a modification to the forces to include a rescaling based on the variation of the velocity coordinates, and a recasting of the evolved quantity as $\mathcal{J}_u f$ instead of $f$, we can utilize a mapped velocity grid to more easily span the expected energy scales in a relativistic plasma. 

To derive an energy conserving DG scheme for this general velocity coordinate kinetic equation, we divide up our phase space domain, $\mathcal{T}$, into phase space cells $K_j$, multiply \eqr{\ref{eq:VlasovSRNonuniform}} by test functions $w$, and integrate by parts to obtain the discrete weak form 
\begin{align}
  \int_{K_j} w \thinspace & \partial_t \left (\mathcal{J}_u f\right )_h \thinspace d\mvec{z} + 
  \oint_{\partial K_j} w^- \mvec{n}\cdot\hat{\mvec{F}}  \thinspace dS \notag \\ 
  & - \int_{K_j} \nabla_{\mvec{z}} w \cdot \gvec{\alpha}_h \left (\mathcal{J}_u f\right )_h \thinspace d\mvec{z} = 0, \label{eq:DGWeakForm}   
\end{align}
where we have abbreviated our advection in phase space to characteristics $\gvec{\alpha}$, and the integration by parts has produced a surface integral term in analogy with finite volume methods and a volume term in analogy with finite element methods.
Note that now the phase space variable $\mvec{z} = (\mvec{x}, \eta^x, \eta^y, \eta^z)$ for each of the (up to) three mapped (four-) velocity coordinates. 
The discrete distribution function $f_h$ is expanded in these same test functions, $w$, 
\begin{align}
    \left (\mathcal{J}_u f\right )_h(\mvec{z}, t) = \sum_{k=1}^{N_p} \left (\mathcal{J}_u f\right )_k(t) w_k(\mvec{z}),
\end{align}
where $N_p$ is the number of test functions which span our solution space and define our complete basis.

In the non-relativistic limit, conservation of energy in the semi-discrete, time-continuous limit is obtained with the proper choice of test functions. 
If the solution space is piecewise polynomials, and we use at least quadratic polynomials in velocity space, then $w = m|\mvec{v}|^2/2$ is in the solution space and can be substituted into \eqr{\ref{eq:DGWeakForm}}. 
Upon summing over all cells, and setting $\mathcal{J}_u = 1$, we obtain
\begin{align}
    \sum_j \int_{K_j} m\frac{|\mvec{v}|^2}{2} & \partial_t f_h + \underbrace{\sum_j \oint_{\partial K_j} m\frac{|\mvec{v}|^2}{2} \mvec{n}\cdot\hat{\mvec{F}}  \thinspace dS}_{=0, \textrm{telescopic sum}} \notag \\ 
    & - \underbrace{\sum_j \int_{K_j} m \mvec{v} \cdot \gvec{\alpha}_h f_h \thinspace d\mvec{z}}_{\sum_k \int_{\Omega_k} \mvec{J}_h \cdot \mvec{E} \thinspace d\mvec{x}} = 0.
\end{align}
Here, we used the fact that $w = m|\mvec{v}|^2/2$ is contained in the solution space and thus continuous at surface interfaces so that a consistent numerical flux, such as central or upwind fluxes, must cancel on either side of the interface. 
Likewise, we have used the definition of the current density to reduce the integration over the phase space to one over the configuration space. 
The total energy of the system can then be conserved in the time continuous limit, or bounded and monotonically decaying, depending on the discretization of Maxwell's equations \citep{Cheng:2014b, Juno:2018, Kormann:2025}.


There are thus two critical components to a DG discretization of the kinetic equation obeying a discrete conservation of energy relation: the surface terms must be a telescopic sum and pairwise cancel when we sum over all cells, and the volume term must set the definition of the electromagnetic work, i.e., whatever the velocity gradient of the discrete energy moment is gives us the definition of the current density. 
If our solution space is piecewise polynomials,
\begin{align}
    \mathcal{V}_h^p = \{ w:w|_{K_j} \in \mvec{P}^p,\forall K_j \in \mathcal{T} \}.
\end{align}
then we can define the continuous subspace,
\begin{align}
    \mathcal{W}_{0,h}^p =  \mathcal{V}_h^p \cap C^0(\mvec{z})
\end{align}
where $\mvec{z}$ is the entire phase-space domain and $C^0(\mvec{z})$ is the set of all continuous functions.
With these definitions, we can then define the discrete representation of the Lorentz factor, $\gamma_h \in \mathcal{W}_{0,h}^p$, and the discrete representation of the relativistic current utilizing the transformed gradient operator,
\begin{align}
    \int_{\Omega_k} \phi \mvec{J}_h \thinspace d\mvec{x} &= \sum_j c^2 \int_{K_j\cap \Omega_k} \phi \left (\nabla_{\gvec{\mathcal{U}}} \gamma_h \right ) \left (\mathcal{J}_u f\right )_h \thinspace d\mvec{z}, \notag \\ 
    & = \sum_j c^2 \int_{K_j\cap \Omega_k} \phi \left (\tdbasis{{i}} \partial_{\eta^i} \gamma_h \right ) \left (\mathcal{J}_u f\right )_h \thinspace d\mvec{z}, \label{eq:relativistic_current}
\end{align}
for all test functions $\phi$ in the configuration-space restriction of the full phase space solution space\footnote{The cells in configuration space are denoted by $\Omega_k\in\mathcal{T}_\Omega$, where $\mathcal{T}_\Omega$ is the restriction of the phase-space mesh $\mathcal{T}$ to configuration space, and we introduce the solution space
$\mathcal{X}^p_{h} = \{ \phi : \phi|_{\Omega_k} \in \mvec{P}^p, \forall \Omega_k \in \mathcal{T}_{\Omega} \}$.
These basis, and test, functions are defined only on the configuration space domain $\Omega$ and thus contain only dependence on the configuration space variable $\mvec{x}$.}.
In fact, in analogy with previous work on Hamiltonian methods \citep{Hakim:2019, Johnson:2026}, noting that the Hamiltonian of a relativistic particle is $H_h = mc^2 \gamma_h$ in the continuous limit, we can then obtain natural discrete representations of the relativistic velocity $(\mvec{u}/\gamma)_h = \nabla_{\gvec{\mathcal{U}}} (H_h/m) = \tdbasis{{i}} \partial_{\eta^i} \gamma_h$ and prove discrete phase space incompressibility and $L^2$ stability. 
We refer the reader to the Methods section for the complete proofs of energy conservation with these definitions of the Lorentz factor and relativistic current, including the mixed nodal-modal form to handle the scaling factors from derivatives in the mapped (four-) velocity space, details on phase space incompressibility and $L^2$ stability, and an empirical demonstration of these properties on a suite of validation tests against the relativistic variants of counter-streaming plasma instabilities.

\subsection{Electric field screening in pair production discharges}

\begin{figure*}
    \centering
    \includegraphics[width=\linewidth]{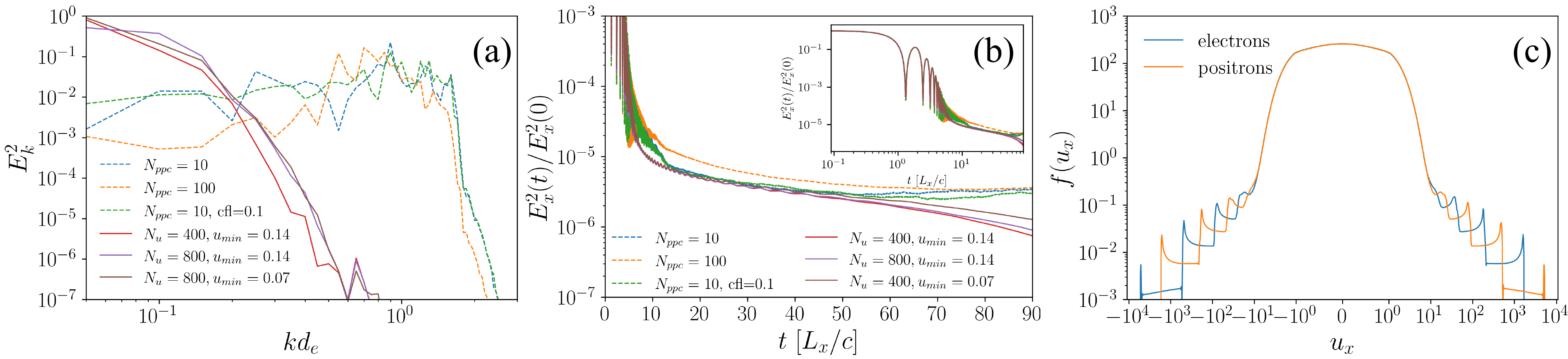}  
    \caption{Electric-field screening by continuously injected electron--positron plasma: comparison between \texttt{Gkeyll} and \texttt{TRISTAN-MP v2} simulations. (a) Spectrum of electric-field fluctuations at late times, highlighting an excess of short-wavelength oscillations in the \texttt{TRISTAN-MP v2} simulations (dashed lines) in comparison to \texttt{Gkeyll} runs (solid lines). (b) Time evolution of the electric-field energy density; the inset highlights the early-time behavior. At late times, \texttt{TRISTAN-MP v2} (dashed lines) exhibits growth of the electric field, whereas \texttt{Gkeyll} (solid lines) shows physically consistent decay. (c) Electron and positron distribution functions, showing peaks at large $u$ that indicate efficient acceleration at early times in the unshielded field, and a broad peak near $u \sim 0$ corresponding to plasma injected at late times after the field becomes marginally screened.
}
    \label{fig:discharge}
\end{figure*}

To explicitly demonstrate the advantages of a grid-based method over a traditional particle-based approach, we compare the two numerical techniques using a benchmark problem inspired by \citet{Tolman:2022}, which analyzes the impact of pair production on the screening of a background electric field. 
The setup is designed to emulate the cyclic production of pair plasma expected to occur in pulsar polar caps \citep{Timokhin:2010, Timokhin:2013}. 
In this environment, a rotation-induced electric field accelerates electrons extracted from the neutron star surface to ultra-relativistic energies. 
These electrons emit high-energy curvature photons that subsequently convert into electron–positron pairs in the presence of a strong magnetic field. 
The resulting plasma loading of the accelerating region progressively shields the electric field component parallel to the magnetic field. Continuous pair injection excites coherent plasma waves, a subset of which are electromagnetic and give rise to coherent radio emission. 

When appropriately scaled to astrophysical parameters, this mechanism reproduces key observational properties of pulsar radio emission \citep{Philippov:2020, Bransgrove:2023, Chernoglazov:2024}, offering a promising pathway toward resolving the long-standing pulsar emission problem. 
Despite this progress, significant numerical challenges remain. 
In particular, fully disentangling the intermittency of pair discharges and the subsequent wave propagation in the pulsar magnetosphere is hindered by the extent to which particle-in-cell (PIC) noise contaminates the electromagnetic emission signal as the density of electron-positron pairs increases. 

In the reference frame co-moving with the ultra-relativistic particle beam, electron-positron pairs are continuously injected into a region with a non-zero electric field. 
To mimic this behavior, we initialize an electron-positron plasma immersed in an almost uniform electric field and continuously load fresh pairs. 
The strength of the electric field is chosen to be sufficient to accelerate particles to a Lorentz factor of $\sim 5 \times 10^3$ over the timescale of the first screening episode, $t\lesssim 1L_x/c$ (see the inset in Fig.~\ref{fig:discharge}b and the sharp peaks at large values of the four-velocity in Fig.~\ref{fig:discharge}c). 
A constant source of non-drifting, uniform-temperature electrons and positrons is injected into the simulation box with an injection rate $\dot{n} = 0.5\, n_0 \omega_{pe}$, such that by the end of the simulation, $t_{\mathrm{end}} = 1{,}800\, \omega_{pe}^{-1}$, the total plasma density has increased by approximately a factor of $10^3$. 
Here $\omega_{pe} = \sqrt{e^2 n_0 / (\epsilon_0 m_e)}$ is the reference electron plasma frequency. 
Both the initial and injected plasma follow a Maxwell-J\"uttner distribution in velocity space, generated using the procedure described in \citet{Johnson:2025}.
A set of simulations is performed with varying velocity-space resolution, using the mapping
\begin{align}
    u_x(\eta^x) = u_{\min}\, \eta^x N_u \pm \exp\!\left(\pm \eta^x \log u_{\max}\right) \mp 1 ,
\end{align}
where $u_{\min}$ sets the minimum resolved four-velocity, $N_u$ is the total number of velocity-space cells, and $u_{\max}$ is the maximum four-velocity. 
The upper and lower signs correspond to the positive and negative velocity half-planes, respectively. 
For comparison, we perform simulations identical to those of \citet{Tolman:2022} using the same PIC code, \texttt{TRISTAN-MP v2} \citep{Spitkovsky:2005, Hakobyan:2024}. 
To distinguish between the impact of Monte Carlo noise and controlled perturbations, both sets of simulations are seeded with five wave modes, $k = 2\pi j / L_x$ for $j = 1,\dots,5$. We set a perturbation amplitude $\alpha = 0.2$, which is applied to both the charge density and the electric field, ensuring that Gauss's law, $\nabla_{\mvec{x}} \cdot \mvec{E} = \rho_c / \epsilon_0$, is satisfied at $t=0$. 
Further details of the simulation setup are included in the Methods section.

We present the results of this suite of simulations in Figure~\ref{fig:discharge}, comparing the electric-field power spectrum averaged over the final hundred fiducial inverse plasma periods, the temporal evolution of the electric-field energy, and the resulting electron and positron distribution functions from the grid-based simulation with $N_u = 800$ and $u_{\min} = 0.14$ to the corresponding \texttt{TRISTAN-MP v2} results. 
We focus on several features of this comparison: the absence of small-scale modes at large $k$ in the \gke\ simulations relative to the \texttt{TRISTAN-MP v2} simulations; the difficulty of shielding these small-scale modes in PIC simulations even with increased particle-per-cell counts and smaller timesteps; the continuous, physically consistent damping of the electric field in the \gke\ runs compared to \texttt{TRISTAN-MP v2}; and the smooth representation of the distribution function over four orders of magnitude in four-velocity enabled by the non-uniform velocity-space mapping. 
We further emphasize that the cost of the grid-based solver is comparable to that of the corresponding PIC run with a fiducial number of particles per cell $N_{\mathrm{ppc}} \approx 1000$. For \gke, the most expensive stage occurs at early times, when particles experience rapid acceleration in the initially unshielded electric field. 
This behavior limits the simulation timestep through the stability condition, $e E \Delta t/m_e < \Delta u$, where $\Delta t$ is the timestep and $\Delta u$ is the (four-) velocity-space cell size. 
In contrast, the PIC simulation evolves rapidly at early times and slows down at later stages, when the overdensity becomes large and the code must advance a large number of simulation particles.

\gke's ability to handle significant dynamic range in density is particularly noteworthy. 
This contrasts with PIC simulations, where increasing the number of particles per cell does not fully suppress the growth of spurious grid-scale modes. 
In particular, previous studies of plasma processes around neutron stars had to limit the pair multiplicity relative to the expected physical values. 
The success of the grid-based solver presented here, however, enables more extensive parameter scans that can probe the detailed mechanisms of electromagnetic emission in these extreme environments.

\subsection{Relativistic magnetic reconnection}

\begin{figure*}
    \centering
    \includegraphics[width=\linewidth]{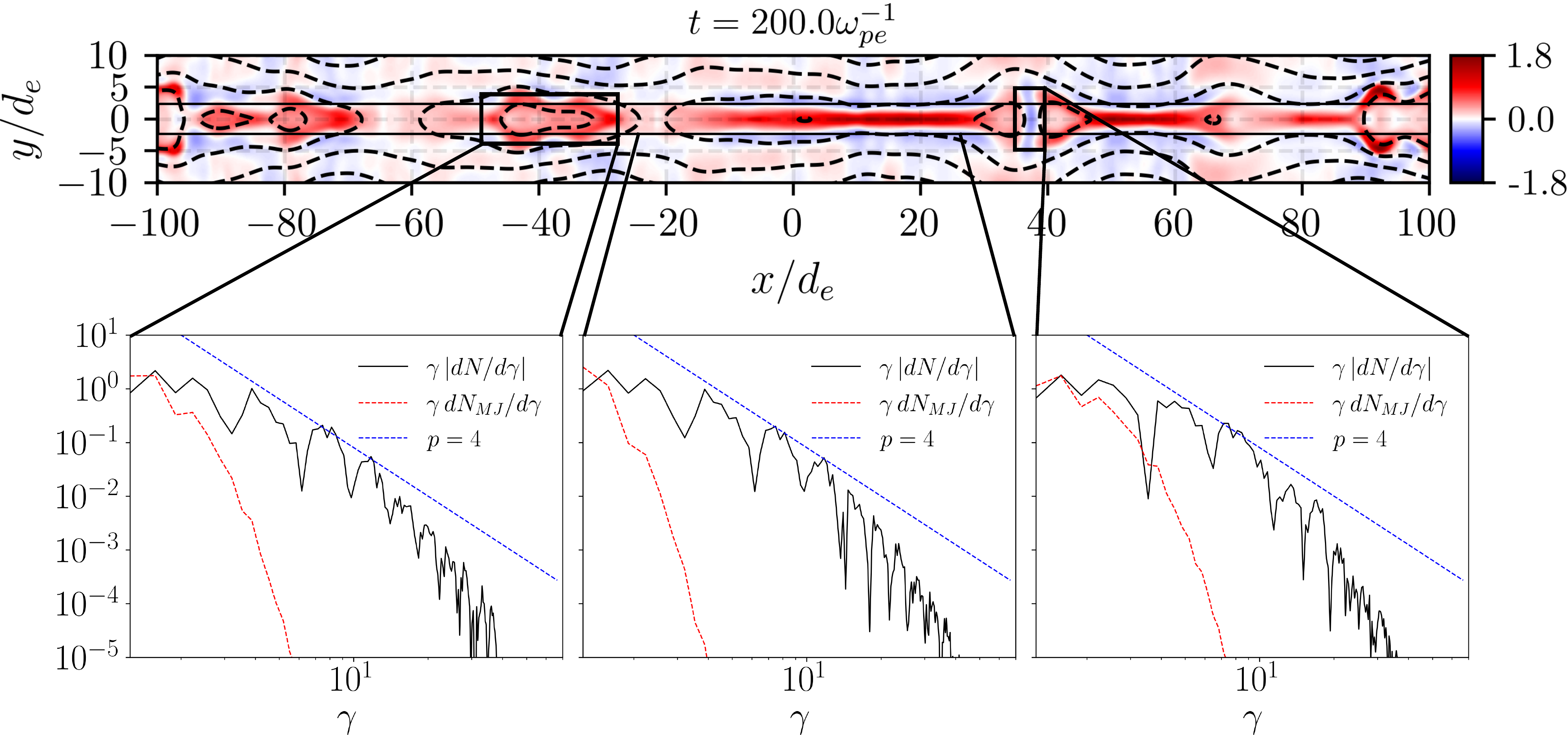}  
    \caption{2x3v \texttt{Gkeyll} simulation of relativistic magnetic reconnection in pair plasma with magnetization parameter $\sigma = 1$. The top row shows the electric-current distribution, while the bottom rows display particle distribution functions sampled from different regions of the current layer. A reference power-law spectrum, $dN/d\gamma \propto \gamma^{-p}$ with $p = 4$, is overplotted to facilitate comparison with previous PIC studies. A Maxwell–Jüttner energy spectrum, constructed using the average plasma-frame density, bulk velocity, and temperature, is also shown to highlight the efficient production of high-energy particles. The \texttt{Gkeyll} energy spectrum is plotted using absolute values to suppress small gaps arising from negative values of the distribution function caused by discretization errors.}
    \label{fig:recon}
\end{figure*}
As a final non-trivial test of the grid-based relativistic Vlasov-Maxwell solver, we perform a simulation of magnetic reconnection in the relativistic regime. 
In this process, magnetic field lines rearrange their topology to reach a lower-energy state, and the released magnetic energy is transferred to the plasma.

At moderate and high magnetization, where the magnetic energy density is comparable or exceeds the plasma rest-mass energy density, this energy conversion does not merely heat the plasma but instead leads to substantial non-thermal particle acceleration \citep{Guo:2014,Werner:2017}. 
Previous PIC studies have shown that relativistic reconnection under these conditions produces robust power-law particle spectra which, at low Lorentz factors $\lesssim \mathrm{few}\times\sigma$, harden with increasing magnetization \citep{Sironi:2014}. 
Here, $\sigma = B^2 / (\mu_0 n m c^2)$ is the plasma magnetization parameter, which sets the characteristic Lorentz factors of particles energized through reconnection. 
Typically, analyses of the distribution function rely on global volume integration to increase particle counts and improve statistical convergence. Here, to demonstrate the utility of a grid-based method, we examine whether the same energetic particle spectra can be recovered using strictly local phase-space diagnostics.

We initialize a force-free current sheet with the electric current equally shared between electrons and positrons at $t=0$, include a small out-of-plane guide field $B_g = 0.01 B_0$, and employ Maxwell--J\"uttner initial plasma conditions. 
Simulations are performed in two spatial and three velocity dimensions (2x3v). 
We utilize linear-to-quadratic mappings in each velocity-space direction,
\begin{align}
    u_i(\eta^i) = u_{\min}\eta^i \frac{N_u}{2} \pm u_{\max}(\eta^i)^2 ,
\end{align}
We choose an initial magnetization $\sigma = B^2 / (\mu_0 n m c^2) = 1$, for which PIC simulations showed a particle spectrum $dN/d\gamma \propto \gamma^{-p}$ with $p=4$ in the zero-guide-field limit \citep{Sironi:2014}. 
Additional details of the force-free configuration are provided in the Methods section.

Figure~\ref{fig:recon} shows the current density evolution with superimposed contours of the in-plane magnetic field at $t = 200\,\omega_{pe}^{-1}$, after reconnection has fully developed.
This time corresponds to approximately one light-crossing time across the elongated dimension of the domain, the characteristic timescale for particle acceleration during reconnection. 
At the same time, we compute electron distribution functions integrated over several distinct spatial subvolumes, binning the distributions as a function of the Lorentz factor $\gamma$. 
The positron evolution is symmetric and exhibits similar behavior.

Across all integration volumes, we observe the emergence of the expected power-law spectrum $dN/d\gamma \propto \gamma^{-4}$ \citep{Sironi:2014}. 
Crucially, this result is obtained without integrating over the full simulation domain to accumulate sufficient statistics. 
The ability to recover nonthermal acceleration locally demonstrates the increased fidelity of the grid-based approach and enables detailed, spatially resolved phase-space analyses.
This capability brings within reach relativistic generalizations of phase-space diagnostics such as the field--particle correlation \citep{Klein:2016,Howes:2017,Klein:2017b,Klein:2020,Horvath:2020,Juno:2021,Juno:2023,Howes:2025} and kinetic pressure-tensor diagnostics \citep{Conley:2024}, representing a particularly exciting new capability enabled by grid-based relativistic Vlasov solvers.
\section{Discussion}\label{sec:discussion}
In this work, we have demonstrated a first-of-its-kind energy-conserving and $L^2$-stable relativistic Vlasov--Maxwell solver based on a class of modal discontinuous Galerkin methods that have previously been successfully employed in non-relativistic plasma modeling. Here, these methods are extended to include not only special relativity, but also general velocity-space grids, enabling the treatment of the vast energy scales expected in high-energy astrophysics and high-energy-density physics.
The solver is implemented in the open-source simulation framework \gke, which, through its existing infrastructure, supports massively parallel simulations on both CPU and GPU architectures.
As representative examples, the magnetic reconnection simulation required $\mathcal{O}(5\times10^5)$ CPU-hours on the Frontera cluster, while the pair-discharge and instability simulations were run on one to four GPUs, depending on resolution.

Beyond the capabilities demonstrated here, we plan further advances in the grid infrastructure and improvements in the accuracy of the numerical scheme. 
The current nonuniform mesh capabilities assume independent mappings in each velocity coordinate; while flexible, this choice is not always optimal for relativistic applications. 
In particular, it may be advantageous to define spherical coordinate systems in velocity space, which can naturally incorporate nonuniform spacing in the total energy, $\gamma$. Additional benefits for problems involving strongly magnetized relativistic plasmas may be obtained by employing velocity-space coordinates aligned parallel and perpendicular to the local magnetic field, as recently demonstrated in \citet{Juno:2025}. In certain relativistic systems, such as the inner magnetospheres of pulsars, strong synchrotron cooling effectively aligns particle velocities with the local magnetic field, which may vary both spatially and temporally. In such cases, the effective dimensionality of the phase space can be reduced from 3x3v to 3x1v, substantially lowering the computational cost. While the present method does not enforce exact charge conservation,
$\nabla_{\mathbf{x}} \cdot \mathbf{E} = \rho_c/\epsilon_0$,
nor strict positivity of the distribution function, $f > 0$, the simulations performed thus far exhibit negligible violations of these constraints. Importantly, these errors do not affect either energy conservation or $L^2$ stability, a property that likely arises from the high-order nature of the DG algorithm employed. Nevertheless, a priori control of these errors is not guaranteed in production simulations, and the incorporation of divergence-cleaning techniques and positivity-preserving limiters therefore remains an important direction for future work.

We can also look ahead to further improvements in the physical fidelity of our numerical framework. 
In particular, fully self-consistent simulations of pair discharges should include the propagation of high-energy photons and their conversion into electron--positron pairs, effects that are neglected in the present work. 
The inclusion of photons can be naturally achieved using the same nonuniform mesh capabilities and Hamiltonian formalism, with the photon Hamiltonian given by $H = |\mathbf{p}|c$, where $|\mathbf{p}|$ is the magnitude of the photon momentum. 
This approach would enable a straightforward coupling between grid-based plasma and photon dynamics through the same set of interactions that has been successfully employed in PIC-based frameworks \citep[e.g.,][]{Chernoglazov:2024}. More broadly, the fact that the nonuniform mesh infrastructure developed here integrates seamlessly with Hamiltonian formulations opens the door to a wide range of future applications. 
Beyond special-relativistic systems, it enables the combination of the advances presented in this work with recent progress on general Hamiltonian systems \citep{Johnson:2026}, paving the way toward a general-relativistic Vlasov solver on nonuniform velocity-space meshes. 
Such a capability would allow the direct incorporation of strong-gravity effects in kinetic simulations of compact objects.

Even prior to the additional advances in numerical flexibility, robustness, and physical fidelity discussed above, the present results already provide a clear path forward for a broad class of relativistic plasma applications.
The solver is currently being deployed in novel physics studies, including the work of \citet{Zeng:2026} on a potential radio emission mechanism from quiescent magnetars. In that scenario, currents driven by global twists of the magnetic field, in combination with radiative cooling, trigger two-stream instabilities that generate electrostatic turbulence, ultimately exciting superluminal electromagnetic modes capable of escaping the magnetar magnetosphere. This study highlights the advantages of accurately capturing even weak electromagnetic signals produced by plasma instabilities. Additional long-term studies of pair-production discharges are currently underway. 
Together, these developments point to a unique opportunity to complement existing kinetic plasma modeling capabilities in the community with a conservative, noise-free approach to solving the relativistic Vlasov--Maxwell system.

\section*{Acknowledgements}
J.~Juno, A.~Philippov, and A.~Hakim gratefully acknowledge the impact of W.~Dorland (1965--2024) on their careers and his facilitating of this collaboration. His insights and energy are missed daily and this paper is dedicated to his memory. 

J.~Juno and A.~Hakim, and the development of \texttt{Gkeyll} were partly funded by the NSF-CSSI program, Award No. 2209471. G.~Johnson was supported by the U.S. Department of Energy, Office of Science, Office of Advanced Scientific Computing Research, Department of Energy Computational Science Graduate Fellowship under Award Number DE-SC0021110. J.~Juno, G.~Johnson, and A.~Hakim were also supported by the U.S. Department of Energy under Contract No. DE-AC02-09CH1146 via LDRD grants. A.~Philippov acknowledges support from the NSF CSSI program (award no.~2311800), the Simons Foundation (grant no.~00001470), an Alfred P.~Sloan Fellowship, and a Packard Foundation Fellowship in Science and Engineering. S.~Zeng and A.~Philippov also acknowledge support by NSF-AST Award No. 2307395.
A.~Chernoglazov was supported by Martin A. and Helen Chooljian Member Fund and the Fund for Memberships in Natural Sciences.
This research is part of the Frontera computing project at the Texas Advanced Computing Center. Frontera is made possible by National Science Foundation (NSF) Award OAC-1818253.

\section*{Data availability}
\gke~ is open source and can be installed by following the instructions on the \gke~ website (\url{http://gkeyll.readthedocs.io}). 
The input file for the \gke~ the simulation presented here is available in the following GitHub repository, \url{https://github.com/ammarhakim/gkyl-paper-inp}.
All plots in this paper utilize the \texttt{postgkyl} package \url{https://github.com/ammarhakim/postgkyl} with dependencies \texttt{matplotlib} \citep{Hunter:2007}, \texttt{numpy} \citep{Harris:2020}, and \texttt{plotly} \url{https://github.com/plotly/plotly.py}.

\section*{Declaration of interests}
The authors report no conflict of interests.

\section*{Declaration of generative AI and AI-assisted technologies in the writing process}
During the copy-editing and LaTeX formatting stage of the writing process, the authors utilized ChatGPT 5.2 to assist with editing and formatting of the TeX document. The authors also acknowledge Codex 5.2 for assistance with debugging the binning procedure to produce the energy spectra in Figure~\ref{fig:recon}, \texttt{plotly}'s API for producing the three-dimensional figures in Figure~\ref{fig:Weibel_evolution}, and \texttt{matplotlib} formatting of the figures in the manuscript. No generative AI was utilized in the design, implementation, and deployment of the special relativistic Vlasov solver in \gke. The authors have reviewed all content produced by generative AI and take full responsibility for the content of this manuscript. 

\section{Methods}\label{sec:methods}
\subsection{Nodal representation of velocity-space Jacobian}
The subsequent proofs of the properties of our discrete scheme utilize a mixed nodal-modal representation, combining a modal representation of the time-dependent distribution function
\begin{align}
    f_h(\mvec{z}, t) = \sum_{k=1}^{N_p} f_k(t) w_k (\mvec{z}), \label{eq:f_modal}
\end{align}
with a nodal representation of the time-independent velocity-space Jacobian 
\begingroup
\fontsize{9}{11}\selectfont
\begin{align}
    & \mathcal{J}_{u_h} = \sum_{\xi = 1}^{N_u} \mathcal{J}_{u_\xi} \ell_{\xi}(\gvec{\eta}) \notag \\ 
    & = \sum_{\xi_{x}=1}^{N_{u_x}} \sum_{\xi_{y}=1}^{N_{u_y}} \sum_{\xi_{z}=1}^{N_{u_z}} \partial_{\eta^1} u_1^{\xi_{x}} \thinspace \partial_{\eta^2} u_2^{\xi_{y}} \thinspace \partial_{\eta^3} u_3^{\xi_{z}} \ell_{\xi_x}(\eta^x)  \ell_{\xi_y}(\eta^y) \ell_{\xi_z}(\eta^z),
\end{align}
\endgroup
where $w_k(\mvec{z})$ are the same modal, orthonormal basis functions used to span the solution space, and we have factorized the nodal representation of the velocity space Jacobian into three independent one-dimensional nodal bases, $\ell_{\xi_i}$.
These nodal bases are Gauss-Legendre nodal bases whose nodal points coincide with the Gauss-Legendre abscissas for an order $p_v+1$ quadrature rule, where $p_v$ is the polynomial order in velocity space. 
For tensor product bases, the modal, orthonormal representation is related to to the nodal Gauss-Legendre basis via the Vandermonde matrix,
\begin{align}
    \mathcal{V}_{jk} = w_k(\xi_j),
\end{align}
In other words, the entries of the matrix $\mathcal{V}_{jk}$ are each of the $k$ modal basis functions evaluated at each of the $j$ Gauss-Legendre abscissas, and one can convert to a nodal representation by taking the modal expansion, such as \eqr{\ref{eq:f_modal}}, and multiplying by the Vandermonde matrix to obtain the $j$-th nodal coefficient \citep{Hesthaven:2007}.

With these definitions and the equivalence between the nodal and modal representations via the Vandermonde matrix, we note that because the velocity-space Jacobian is a time-independent quantity and represented nodally, we know the velocity-space Jacobian point-wise at Gauss-Legendre abscissas. 
With this point-wise information, the modal representation of $(\mathcal{J}_u f)_h$ can be constructed at $t=0$ using projection onto our modal basis at Gauss-Legendre quadrature points, including the additional nodal $\mathcal{J}_u$ factor; for example, we can utilize the method of projecting the relativistic generalization of the Maxwellian distribution function onto our basis described in \cite{Johnson:2025} with a multiplication at each velocity-space Gauss-Legendre quadrature point by the evaluation of the velocity-space Jacobian at that same velocity-space quadrature point. 
In this way, we can obtain a uniformly high-order representation of the solution including the geometry. 
This representation allows us to avoid the simplification of linear mapping and piecewise constant velocity-space Jacobian used in \citet{Francisquez:2026}, which may reduce the order of the solution when the mapping is highly nonuniform in each velocity-space dimension.

This point-wise information is then a critical component when examining individual pieces of the update,
\begingroup
\fontsize{8}{10}\selectfont
\begin{align}
    -& \int_{K_j}  \left (\partial_{u_x} w \right )\alpha_{u_x} f_h \thinspace d\mvec{z} = -\int_{K_j} \frac{\left (\partial_{\eta^1} w \right )}{\left (\partial_{\eta^1} u_1 \right )}\left (\mathcal{J}_u f \right )_h \notag \\
    & \left ( \frac{q_s}{m_s}\left [E_{x_h} + \frac{\left (\partial_{\eta^2} \gamma_h \right )}{\left (\partial_{\eta^2} u_2 \right )} B_{z_h} - \frac{\left (\partial_{\eta^3} \gamma_h \right )}{\left (\partial_{\eta^3} u_3 \right )} B_{y_h} \right ] \right ) \thinspace d\mvec{z}.
\end{align}
\endgroup
which contain non-polynomial expressions from the inverse Jacobian component factors that we wish to eliminate using the point-wise information we have. 
To perform this simplification though, we must be able to convert between the nodal and modal representations, even in this nonlinear product. 
In general, this conversion is not possible; the nonlinear products are not contained in our basis and therefore any conversion between nodal and modal representations incurs aliasing errors. 
However, for the case when $p_v = 2$, then the nonlinear products in the degrees of freedom we are also dividing by $\partial_{\eta^i} u_i$ are at most order 5 since the Hamiltonian is quadratic, $(\mathcal{J}_u f)_h$ is quadratic, and the basis functions are quadratic, but there is one derivative lowering the order of one of the terms. 
Since Gauss-Legendre quadrature integrates polynomials of order $2N-1$, where $N$ is the number of quadrature points, and we have $p_v + 1 = 3$ quadrature points, then we can safely convert to the nodal representation, divide by the point-wise velocity-space Jacobian information, and convert back to the modal representation for efficient integration:
\begin{align}
-& \int_{K_j} \frac{\left (\partial_{\eta^1} w \right )}{\left (\partial_{\eta^1} u_1 \right )} \frac{q_s}{m_s} E_{x_h} \left (\mathcal{J}_u f \right )_h \thinspace d\mvec{z} = \notag \\
& -\int_{K_j} \left (\partial_{\eta^1} w \right ) \frac{q_s}{m_s} E_{x_h}  \left (\partial_{\eta^2} u_2 \thinspace \partial_{\eta^3} u_3 \thinspace f \right )_h \thinspace d\mvec{z}, \label{eq:EfieldJacCancel} \\ 
-& \int_{K_j} \frac{\left (\partial_{\eta^1} w \right )}{\left (\partial_{\eta^1} u_1 \right )} \frac{q_s}{m_s} \frac{\left (\partial_{\eta^2} \gamma_h \right )}{\left (\partial_{\eta^2} u_2 \right )} B_{z_h} \left (\mathcal{J}_u f \right )_h \thinspace d\mvec{z} = \notag \\ 
&-\int_{K_j} \left (\partial_{\eta^1} w \right ) \frac{q_s}{m_s} \left (\partial_{\eta^2} \gamma_h \right ) B_{z_h} \left (\partial_{\eta^3} u_3 \thinspace  f \right )_h  \thinspace d\mvec{z}, \label{eq:BzfieldJacCancel} \\
& \int_{K_j} \frac{\left (\partial_{\eta^1} w \right )}{\left (\partial_{\eta^1} u_1 \right )} \frac{q_s}{m_s} \frac{\left (\partial_{\eta^3} \gamma_h \right )}{\left (\partial_{\eta^3} u_3 \right )} B_{y_h} \left (\mathcal{J}_u f \right )_h \thinspace d\mvec{z} = \notag \\ 
& \int_{K_j} \left (\partial_{\eta^1} w \right ) \frac{q_s}{m_s} \left (\partial_{\eta^3} \gamma_h \right ) B_{y_h} \left (\partial_{\eta^2} u_2 \thinspace f \right)_h \thinspace d\mvec{z} \label{eq:ByfieldJacCancel}.
\end{align}
These manipulations can also be utilized to eliminate the non-polynomial divisions in our definitions of the velocity moments, such as the current density, e.g., for the $x$ current
\begingroup
\fontsize{9}{11}\selectfont
\begin{align}
     \int_{\Omega_k} \phi J_x \thinspace d\mvec{x} & = \sum_j c^2 \int_{K_j\cap \Omega_k} \phi \frac{\left (\partial_{\eta^1} \gamma_h \right )}{\left (\partial_{\eta^1} u_1 \right )} \left (\mathcal{J}_u f\right )_h \thinspace d\mvec{z} \notag \\ 
     & = \sum_j c^2 \int_{K_j\cap \Omega_k} \phi \left (\partial_{\eta^1} \gamma_h \right ) \left (\partial_{\eta^2} u_2 \thinspace \partial_{\eta^3} u_3 \thinspace f \right )_h \thinspace d\mvec{z}    
\end{align}
\endgroup

\subsection{Proof of energy conservation}
Starting with \eqr{\ref{eq:DGWeakForm}} and substituting in $w = H_h = mc^2\gamma_h$, and summing over all cells, we obtain,
\begin{align}
  \sum_j mc^2 & \int_{K_j} \gamma_h \partial_t \left (\mathcal{J}_u f\right )_h \thinspace d\mvec{z} + 
  \underbrace{\sum_j mc^2 \oint_{\partial K_j} \gamma_h^- \mvec{n}\cdot\hat{\mvec{F}}  \thinspace dS}_{=0, \textrm{telescopic sum}} \notag \\
  & - \sum_j mc^2 \int_{K_j} \underbrace{\nabla_{\mvec{z}} \gamma_h}_{\tdbasis{{i}} \partial_{\eta^i} \gamma_h} \cdot \gvec{\alpha}_h \left (\mathcal{J}_u f\right )_h \thinspace d\mvec{z} = 0,    
\end{align}
where we have used the fact that $\gamma_h \in \mathcal{W}_{0,h}^p$ and is thus continuous at cell interfaces, along with a choice of consistent flux function such as central or upwind fluxes, to eliminate the surface contributions to the sum.
Each surface flux pairwise cancels because the test function is continuous and the flux is consistent. 
Note the simplification of $\nabla_{\mvec{z}} \gamma_h = \tdbasis{{i}} \partial_{\eta^i} \gamma_h$; $\gamma_h$ is only a function of momentum space. 
We can then utilize \eqr{\ref{eq:relativistic_current}}, along with the definition of the relativistic energy density,
\begin{align}
    \int_{\Omega_k} \phi \mathcal{E}_h \thinspace d\mvec{x} = \sum_j mc^2 \int_{K_j\cap \Omega_k} \phi \gamma_h \left (\mathcal{J}_u f\right )_h \thinspace d\mvec{z}, 
\end{align}
to simplify the volume term
\begin{align}
  \sum_k \int_{\Omega_k} \pfrac{\mathcal{E}_h}{t} \thinspace d\mvec{x} - \sum_k \int_{\Omega_k} \mvec{J}_h \cdot \mvec{E}_h \thinspace d\mvec{x} = 0, \label{eq:SR_energy_conservation}
\end{align}
using the definition of $\gvec{\alpha}_h$ and properties of the cross product to cancel the magnetic field contribution to the total energy evolution. 
At this point, with this definition of the relativistic current density, the remainder of the proof is identical to \cite{Juno:2018} and conservation of energy can be shown utilizing central fluxes for Maxwell's equations, while energy will monotonically decay when utilizing upwind fluxes for Maxwell's equations. 
We emphasize the substitution in \eqr{\ref{eq:SR_energy_conservation}} relies on the accurate integration of the nonlinear production $\mvec{J}_h \cdot \mvec{E}_h$ and no aliasing errors can be tolerated in the integration of this term, as previously discussed in \cite{HakimJuno:2020}. 
\subsection{Proof of phase-space incompressibility}
Phase-space incompressibility in the continuous limit is defined as
\begin{align}
\nabla_{\mvec{z}} \cdot \left (\mathcal{J}_u \gvec{\alpha}_h \right ) = 0.
\end{align}
Even with mapped coordinates in velocity space, the configuration space components of this divergence still trivially vanish in the discrete limit, as $\gamma_h$ has no configuration space dependence and therefore,
\begin{align}
    \nabla_{\mvec{x}} \cdot \left (\mathcal{J}_u \nabla_{\gvec{\mathcal{U}}} \gamma_h \right ) = 0. \label{eq:strongXIncompress}
\end{align}
With the mixed nodal-modal representation, the velocity space divergence similarly vanishes.
Consider the electric field force dropping factors of $q_s/m_s$ for notational brevity:
\begin{align}
    \int_{K_j} w \nabla_{\gvec{\mathcal{U}}} \cdot \left (\mathcal{J}_u \mvec{E}_h \right ) & \thinspace d\mvec{z} = \oint_{\partial K_j} w^- \mvec{E}_h \cdot \hat{\mvec{n}} \mathcal{J}_u^S \thinspace dS \notag \\ 
    & - \int_{K_j} \left (\tdbasis{{i}} \partial_{\eta^i} w \right ) \cdot \left (\mathcal{J}_u \mvec{E}_h \right ) \thinspace d\mvec{z},
\end{align}
where $\mathcal{J}_u^S$ is the surface velocity-space Jacobian, e.g., at the $u_x$ surface it is $\partial_{\eta^2} u_2 \thinspace \partial_{\eta^3} u_3$.
We note that in an identical fashion to \eqr{\ref{eq:EfieldJacCancel}}, the volume term can be manipulated component by component. 
For example, the $u_x$ update becomes
\begin{align}
    \int_{K_j} \frac{\left (\partial_{\eta^1} w \right )}{\left (\partial_{\eta^1} u_1 \right )} \mathcal{J}_u E_{x_h} & = \int_{K_j} \partial_{\eta^1} w \left ( \partial_{\eta^2} u_2 \thinspace \partial_{\eta^3} u_3 \right ) E_{x_h} \notag \\ 
    & = \oint_{\partial K_j} w^- \left (\partial_{\eta^2} u_2 \thinspace \partial_{\eta^3} u_3 E_{x_h} \right )^- \thinspace dS, \label{eq:EfieldIncompress}
\end{align}
because $\partial_{\eta^2} u_2 \thinspace \partial_{\eta^3} u_3 \thinspace E_{x_h}$ has no $\eta^1$ dependence. And since $\partial_{\eta^2} u_2 \thinspace \partial_{\eta^3} u_3 \thinspace E_{x_h}$ has no $\eta^1$ dependence, it is continuous at $u_x$ surfaces, so the two surface terms vanish and component by component we have
\begin{align}
    \int_{K_j} w \nabla_{\gvec{\mathcal{U}}} \cdot \left (\mathcal{J}_u \mvec{E}_h \right ) \thinspace d\mvec{z} = 0.
\end{align}

The magnetic field force can similarly be shown to be incompressible in phase space, but now we must utilize commutativity of partial derivatives $\partial_{\eta^1} \partial_{\eta^2} \gamma_h - \partial_{\eta^2} \partial_{\eta^1} \gamma_h = 0$. 
Performing similar simplifications to \eqr{\ref{eq:BzfieldJacCancel}} and \eqr{\ref{eq:ByfieldJacCancel}}, we obtain for the volume term
\begin{align}
    \int_{K_j} \left (\tdbasis{{i}} \partial_{\eta^i} w \right ) \cdot \left (\tdbasis{{i}} \partial_{\eta^i} \gamma_h \times \mvec{B}_h \mathcal{J}_u \right ) \thinspace d\mvec{z} & = \notag \\ 
    \int_{K_j} \left (\partial_{\eta^1} w \right ) \left [  \underbrace{\left (\partial_{\eta^2} \gamma_h \right ) B_{z_h} \partial_{\eta^3} u_3}_{I} - \underbrace{\left (\partial_{\eta^3} \gamma_h \right ) B_{y_h} \partial_{\eta^2} u_2}_{II} \right ] \thinspace d\mvec{z} \notag \\ 
     + \int_{K_j} \left (\partial_{\eta^2} w \right ) \left [ \underbrace{\left (\partial_{\eta^3} \gamma_h \right ) B_{x_h} \partial_{\eta^1} u_1}_{III} - \underbrace{\left (\partial_{\eta^1} \gamma_h \right ) B_{z_h} \partial_{\eta^3} u_3}_{I} \right ] \thinspace d\mvec{z} \notag \\ 
     + \int_{K_j} \left (\partial_{\eta^3} w \right ) \left [ \underbrace{\left (\partial_{\eta^1} \gamma_h \right ) B_{y_h} \partial_{\eta^2} u_2}_{II} - \underbrace{\left (\partial_{\eta^2} \gamma_h \right ) B_{x_h} \partial_{\eta^1} u_1}_{III} \right ] \thinspace d\mvec{z},
\end{align}
which upon integration by parts in a similar fashion to the manipulations in \eqr{\ref{eq:EfieldIncompress}}, we are left with common factors multiplying mixed partials of the discrete Lorentz boost factor we have marked as terms $I, II, \& \thinspace III$. 
The remaining surface terms then vanish component by component, surface by surface, because $\gamma_h \in \mathcal{W}_{0,h}^p$ and the magnetic field force introduces only transverse derivatives to the surface. 

We thus have in the discrete limit
\begin{align}
    \int_{K_j} w \nabla_{\mvec{z}} \cdot \left (\mathcal{J}_u \gvec{\alpha}_h \right ) \thinspace d\mvec{z} = 0.
\end{align}
We note that, while phase-space incompressibility still holds point-wise, or strongly, for the advection in configuration space, incompressibility holds only weakly in velocity space due to the introduction of the mapping and velocity-space Jacobian. 
But there is a further simplification of this weak phase space incompressibility in velocity-space
\begin{align}
    \int_{K_j} w \partial_{\eta^i} \left (\mathcal{J}^{S_i}_u \gvec{\sigma}_i \cdot \gvec{\alpha}_h \right ) \thinspace d\mvec{z} = 0, \label{eq:weakVIncompress}
\end{align}
where we have used the fact that our mixed nodal-modal formalism allows us to re-express $\tdbasis{i} J_u = J_u^{S_i}$ by canceling the component of the velocity-space Jacobian normal to the derivative. 
It is this form in \eqr{\ref{eq:weakVIncompress}} which is needed for the subsequent $L^2$ stability proof. 
We emphasize that \eqr{\ref{eq:weakVIncompress}} is a novel demonstration of phase space incompressibility even when using mapped coordinates, as previous generalizations of discretizations of the kinetic equation cannot, in general, show phase space incompressibility for arbitrary Jacobian factors \citep{Johnson:2026}.

\subsection{Proof of $L^2$ stability}
We start again with \eqr{\ref{eq:DGWeakForm}}, but now substituting in $w = f_h$, and then summing over all cells to obtain
\begin{align}
  \sum_j &\int_{K_j} f_h \partial_t \left (\mathcal{J}_u f\right )_h \thinspace d\mvec{z} + 
  \sum_j \oint_{\partial K_j} f_h^- \mvec{n} \cdot \widehat{\mvec{F}} \thinspace J_u^{S_i} dS \notag \\ 
  &- \sum_j \int_{K_j} \nabla_{\mvec{z}} f_h \cdot \gvec{\alpha}_h \left (\mathcal{J}_u f\right )_h \thinspace d\mvec{z} = 0,
\end{align}
where we have separated the configuration-space and velocity-space fluxes to make clear how the normal component of the Jacobian factor cancels to leave just the transverse Jacobian, 
\begin{align}
    \widehat{\mvec{F}} & = \widehat{\gvec{\alpha}^S_h f_h}, \\
    \gvec{\alpha}^S_h & = \left ( \gvec{\sigma}_i \left [ \partial_{\eta^i} \gamma_h \right ], \frac{q_s}{m_s} \left [\mvec{E}_h + \tdbasis{{i}} \partial_{\eta^i} \gamma_h \times \mvec{B}_h \right ] \right ). 
\end{align}
And we note again that $\widehat{\mvec{F}}$ defines a consistent numerical flux, such as central fluxes or upwind fluxes.
We now examine the volume term and use the following property of weak equality 
\begin{align}
     \int_{K_j} w_h (f g)_h d\mvec{z} = \int_{K_j} w_h f_h g_h d\mvec{z}, \label{eqn:weak_id_2}
\end{align} 
to write the volume term as 
\begin{align}
    \int_{K_j} \nabla_{\mvec{z}} f_h \cdot \gvec{\alpha}_h \left (\mathcal{J}_u f\right )_h \thinspace d\mvec{z} & = \int_{K_j} \left (\left [\nabla_{\mvec{z}} f \right ] J_u f \right )_h \cdot \gvec{\alpha}_h \thinspace d\mvec{z} \notag \\ 
    & = \int_{K_j} \frac{1}{2} \nabla_{\mvec{z}} J_u f^2_h \cdot \gvec{\alpha}_h \thinspace d\mvec{z},
\end{align}
where we have used our mixed nodal-modal representation to simplify the weak product $\left (\left [\nabla_{\mvec{z}} f \right ] J_u f \right )_h$ at Gauss-Legendre quadrature points where the values of the velocity-space Jacobian are known. 
In fact, there is a further simplification for the velocity-space gradient since $\nabla_{\mvec{z}} f = \tdbasis{i} \partial_{\eta^i} f$ and thus at Gauss-Legendre quadrature points there is a cancellation of the normal component of the velocity-space Jacobian such that 
\begin{align}
    \left (\left [\nabla_{\gvec{\mathcal{U}}} f \right ] J_u f \right )_h & = \left (\left [\tdbasis{i} \partial_{\eta^i} f \right ] J_u f \right )_h \notag \\ 
    & = \left (\left [\partial_{\eta^i} f \right ] J^{S_i}_u \gvec{\sigma}_i f \right )_h \notag \\ 
    & = \frac{1}{2} \left (J^{S_i}_u \gvec{\sigma}_i \left [\partial_{\eta^i} f^2 \right ] \right )_h
\end{align}
since $J_u^S$, the surface velocity Jacobian, has no variation in the direction $\partial_{\eta^i}$ as we showed in the phase-space incompressibility proof. 
For example, at the $u_x$ surface $\partial_{\eta^2} u_2 \thinspace \partial_{\eta^3} u_3$ has no $\eta^1$ dependence. 
With this simplification, we can integrate by parts again and utilize both \eqr{\ref{eq:strongXIncompress}} and \eqr{\ref{eq:weakVIncompress}} to eliminate the volume term and obtain
\begin{align}
  \sum_j & \frac{1}{2} \int_{K_j} \partial_t \left (\mathcal{J}_u f^2 \right )_h \thinspace d\mvec{z} \notag \\ 
  & + 
  \sum_j \oint_{\partial K_j} f_h^- \mvec{n} \cdot \left ( \widehat{\mvec{F}} - \frac{f_h^-}{2} \left (\gvec{\alpha}_h^S \right )^- \right ) \thinspace J_u^{S_i} dS = 0. 
\end{align}
We can make one further simplification to this expression utilizing the fact that $\gvec{\alpha}_h^S$ is continuous at the surface interfaces, $\left (\gvec{\alpha}_h^S \right )^- = \gvec{\alpha}_h^S$ so that 
\begin{align}
  \sum_j & \frac{1}{2} \int_{K_j} \partial_t \left (\mathcal{J}_u f^2 \right )_h \thinspace d\mvec{z} \notag \\ 
  & + 
  \sum_j \oint_{\partial K_j} f_h^- \mvec{n} \cdot \gvec{\alpha}_h^S \left ( \widehat{f} - \frac{f_h^-}{2} \right ) \thinspace J_u^{S_i} dS = 0, 
\end{align}
where $\widehat{f}$ is determined by your choice of flux function, e.g., central
\begin{align}
    \widehat{f} = \frac{1}{2} \left (f_h^- + f_h^+ \right)
\end{align}
or upwind
\begin{align}
    \widehat{f} =     \begin{cases}
        f^- \quad \textrm{if} \quad \sign(\gvec{\alpha}^S_h) > 0, \\
        f^+ \quad \textrm{if} \quad \sign(\gvec{\alpha}^S_h) < 0.
    \end{cases}
\end{align}
We can then obtain the same result as other studies of DG for the kinetic equation, such as \cite{Juno:2018} for the Vlasov equation and \cite{Johnson:2026} for the Boltzmann equation on smooth manifolds that the $L^2$ energy is conserved for central fluxes,
\begin{align}
  \sum_j \frac{1}{2} \int_{K_j} \partial_t \left (\mathcal{J}_u f^2 \right )_h \thinspace d\mvec{z} = 0, 
\end{align}
and monotonically decaying for upwind fluxes
\begin{align}
  \sum_j & \frac{1}{2} \int_{K_j} \partial_t \left (\mathcal{J}_u f^2 \right )_h \thinspace d\mvec{z} = \notag \\  
  & -\frac{1}{4} \sum_j \oint_{\partial K_j} \left | \mvec{n} \cdot \gvec{\alpha}_h^S \right | \left ( f^-_h - f^+_h \right )^2 \thinspace J_u^{S_i} dS \leq 0. 
\end{align}
We emphasize that, similar to the phase space incompressibility proof, this result is a unique demonstration of $L^2$ stability utilizing mapped coordinates, as $L^2$ stability cannot be proved for arbitrary coordinate systems where the Jacobian for the transformation is a general function of the phase space \citep{Johnson:2026}.

\subsection{Validation against linear theory for relativistic streaming instabilities and empirical demonstration of energy conservation and $L^2$ stability} \label{sec:linear}
To both validate the solver against known analytic solutions and empirically demonstrate the proofs of energy conservation and $L^2$ stability, simulate relativistic variants of counter-streaming instabilities in collisionless plasmas, arising from perturbations both parallel and perpendicular to the direction of motion of the drifting plasma \citep{Bret:2009,Bret:2010}. 
Parallel perturbations excite the two-stream \citep{Bohm:1949,Buneman:1958} instability, while perturbations perpendicular to the drift excite the filamentation, or Weibel, instability \citep{Fried:1959,Weibel:1959}.
We consider two counter-streaming electron populations of the form
\begin{align}
    f_e(t=0) \propto & \exp \left (-\frac{\Gamma}{\theta} \left [\gamma - v_b \cdot u_x \right ] \right ) \notag \\ 
    + & \exp \left (-\frac{\Gamma}{\theta} \left [\gamma + v_b \cdot u_x \right ] \right )
\end{align}
for the two-stream instability case, and 
\begin{align}
    f_e(t=0) \propto & \exp \left (-\frac{\Gamma}{\theta} \left [\gamma - v_b \cdot u_y \right ] \right ) \notag \\ 
    + & \exp \left (-\frac{\Gamma}{\theta} \left [\gamma + v_b \cdot u_y \right ] \right )
\end{align}
for the filamentation instability case.  Ions are modeled as an immobile neutralizing background,
For the two-stream instability, we employ a 1x1v simulation geometry, consisting of one spatial dimension and one velocity dimension. In this configuration, the $x$-component of the four-velocity, $u_x$, couples to the electric field $E_x$. For the filamentation instability, we use a 1x2v geometry with one spatial dimension and two velocity dimensions, $u_x$ and $u_y$, which couple self-consistently to $E_x$, $E_y$, and $B_z$. Note that as a result of treating the ions as immobile, the current density in Maxwell’s equations arises solely from electron motion. 
Here, $\Gamma = 1/\sqrt{1 - v_b^2/c^2}$ is the bulk Lorentz factor, $\theta = T/m_e c^2$ is the normalized temperature, and as before $\gamma = \sqrt{1 + |\mvec{u}|^2/c^2}$ is the particle Lorentz factor. 

The counter-streaming populations are initialized with equal densities and temperatures, $n_0 = 0.5$, $T = 0.04$, with the normalizations, $\epsilon_0 = \mu_0 = 1$, $m_e = 1$, and $q_e = -1$. The initial drifts are chosen to be relativistic, with $v_b = 0.99c$ for the two-stream simulations and $v_b = 0.9c$ for the filamentation simulations. 
To resolve the counter-streaming beams while allowing for the generation of energetic particles, we initialize linear-to-quadratic mappings in each velocity-space direction,
\begin{align}
    u_i(\eta^i) = u_{\min}\eta^i \frac{N_u}{2} \pm u_{\max}(\eta^i)^2.
\end{align}
We also run a set of simulations with uniform resolution for the comparison to linear theory. 
For electrostatic waves in a one-dimensional, unmagnetized plasma, linear theory yields the following dispersion relation:

\begin{align}
    1 + \frac{\omega_p^2}{\omega^2} 
    \bigg [
        \int d\mvec{u}  \frac{u_x \partial_{u_x} f_0}{\gamma( 1 - \frac{k_x u_x}{\gamma \omega} ) }
    \bigg ]
    =
    0. \label{eqn:TS_linear_dispersion}
\end{align}
where $\omega$ is the wave frequency, $\omega_p$ is the plasma frequency, $f_0$ is normalized such that its integral over all momentum space equals unity, and $k_x$ is the perturbation wavenumber. For the filamentation case, considering a perturbation of the $B_z$ magnetic-field component propagating along the $x$-direction, $k_x$, with a background plasma drift along the $y$-direction, the resulting linear dispersion relation for the warm relativistic Weibel instability is

\begin{align}
    1 & - \frac{\omega^2}{c^2 k_x^2} - \frac{\omega_p^2}{c^2 k_x^2}
    \int d\mvec{u} \frac{u_y}{\gamma \omega - k_x u_x}
    \left [
        \left(\omega - \frac{k_x u_x}{\gamma}\right)\partial_{u_x} f_0
    \right ] 
    \notag \\ 
    & - \frac{\omega_p^2}{c^2 k_x^2}
    \int d\mvec{u} \frac{u_y}{\gamma \omega - k_x u_x}
    \left [
        \frac{k_x u_x}{\gamma} \partial_{u_y} f_0
    \right ]
    =
    0. \label{eqn:Weibel_linear_dispersion}
\end{align}
Unstable roots of Eqs.~\ref{eqn:TS_linear_dispersion} and~\ref{eqn:Weibel_linear_dispersion} are found numerically by locating the intersections of the zero contours of the real and imaginary parts of the dispersion relation for $\mathrm{Im}(\omega) > 0$, which correspond to unstable solutions.

\begin{figure*}
\centering
\includegraphics[width = 0.9\linewidth]{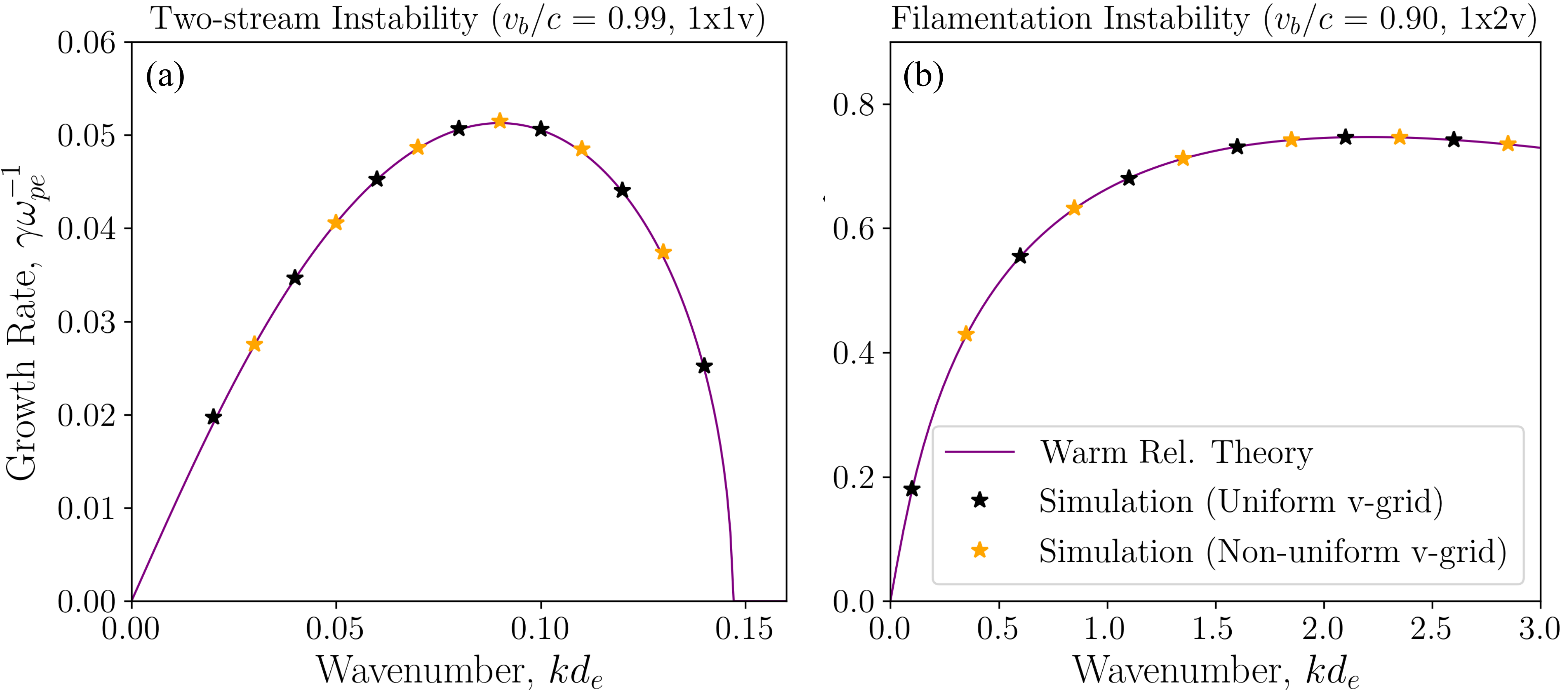}
\caption{Comparison of growth rates of the two-stream (a) and filamentation (b) instabilities between simulations and warm relativistic linear theory as a function of the unstable-mode wavenumber. In both cases, the fluid-frame moments of the initial distribution are $n_0 = 1.0$ (the sum of the densities of both beams) and $T_0/m_0c^2 = 0.04$. The initial beam velocities are $v_b/c = 0.99$ for the two-stream case and $v_b/c = 0.9$ for the filamentation case. Simulations are performed using both uniform and non-uniform grids in velocity space.
}\label{fig:TS_and_Weibel_simulation_vs_theory}
\end{figure*}

We scan wavenumbers in our \texttt{Gkeyll} simulations by setting the box size to $L_x = 2\pi/k_x$ and perturbing the largest mode in the box. For the two-stream instability, this perturbation is applied to the electron density along with the corresponding electric-field fluctuation, while for the filamentation instability, we perturb the $B_z$ component of the magnetic field. Figure~\ref{fig:TS_and_Weibel_simulation_vs_theory} compares the linear theory predictions with numerical simulations across a range of wavenumbers for both instabilities. The resolution is fixed at $N_x \times N_v = 128 \times 128$ for the two-stream simulations and $N_x \times N_v^2 = 96 \times 48^2$ for the filamentation simulations. All simulations employ a tensor-product quadratic polynomial basis.

\begin{figure*}
\centering
\includegraphics[width = 0.9\linewidth]{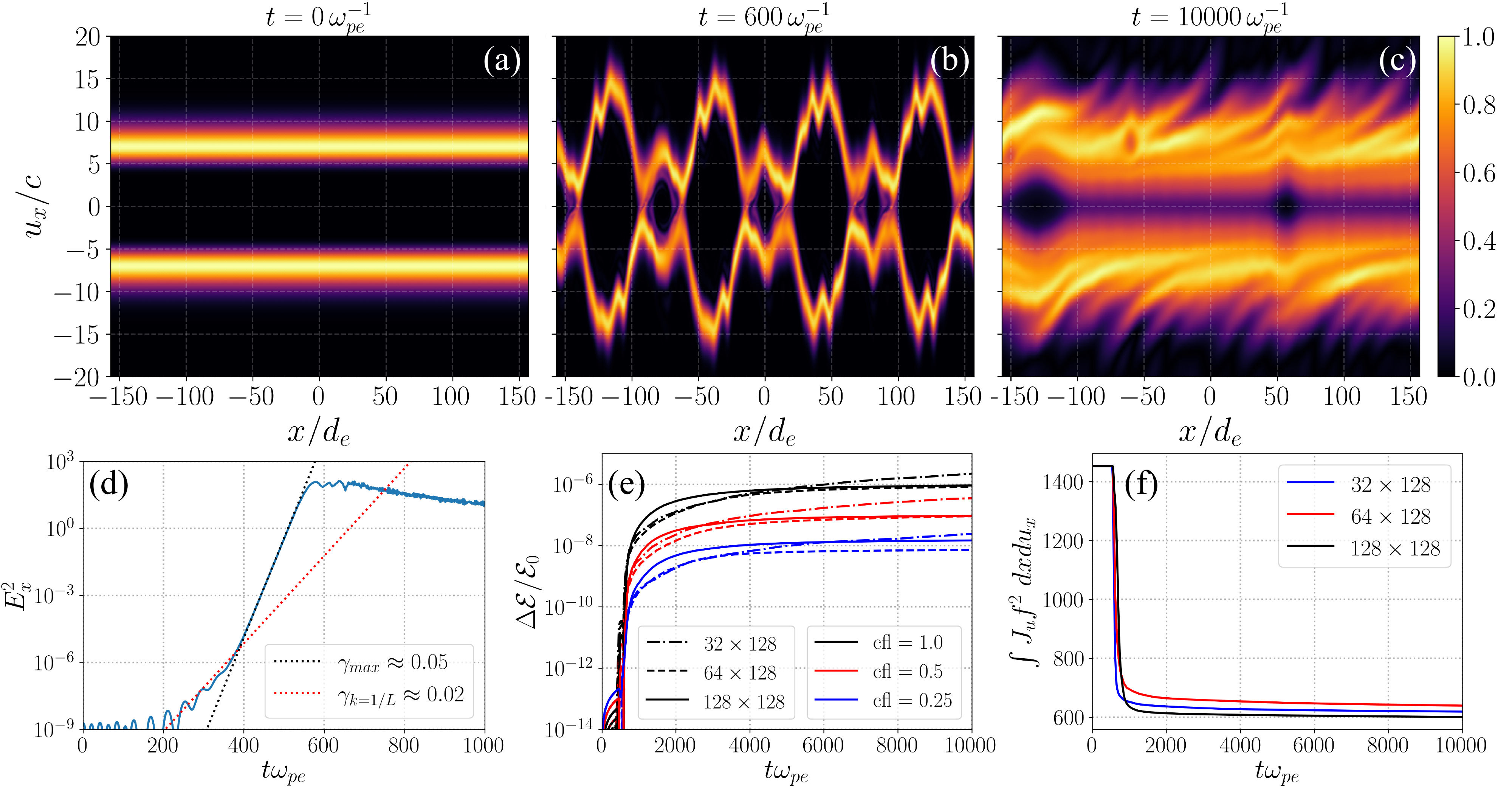}
\caption{Evolution of the two-stream instability from initial conditions (a) into the initial nonlinear saturation (b) and long-time evolution (c) in phase space. The distribution function is normalized to its peak value at each time for clarity of visualization. Panel (d) shows the evolution of the energy in the electrostatic, $E_x$, component. We observe an initially slower-growing box-scale mode, followed by the emergence and dominance of the fastest-growing mode in the system. The relative change in total energy (e) demonstrates that energy conservation is governed by the time-step size, exhibiting convergence consistent with $(\Delta t)^3$, as expected for the third-order strong-stability-preserving Runge–Kutta method used for temporal discretization. Panel (f) shows the integrated $L^2$ energy, which decreases monotonically across resolutions, indicating $L^2$ stability.}\label{fig:TS_evolution}
\end{figure*}

\begin{figure*}
\centering
\includegraphics[width = 0.9\linewidth]{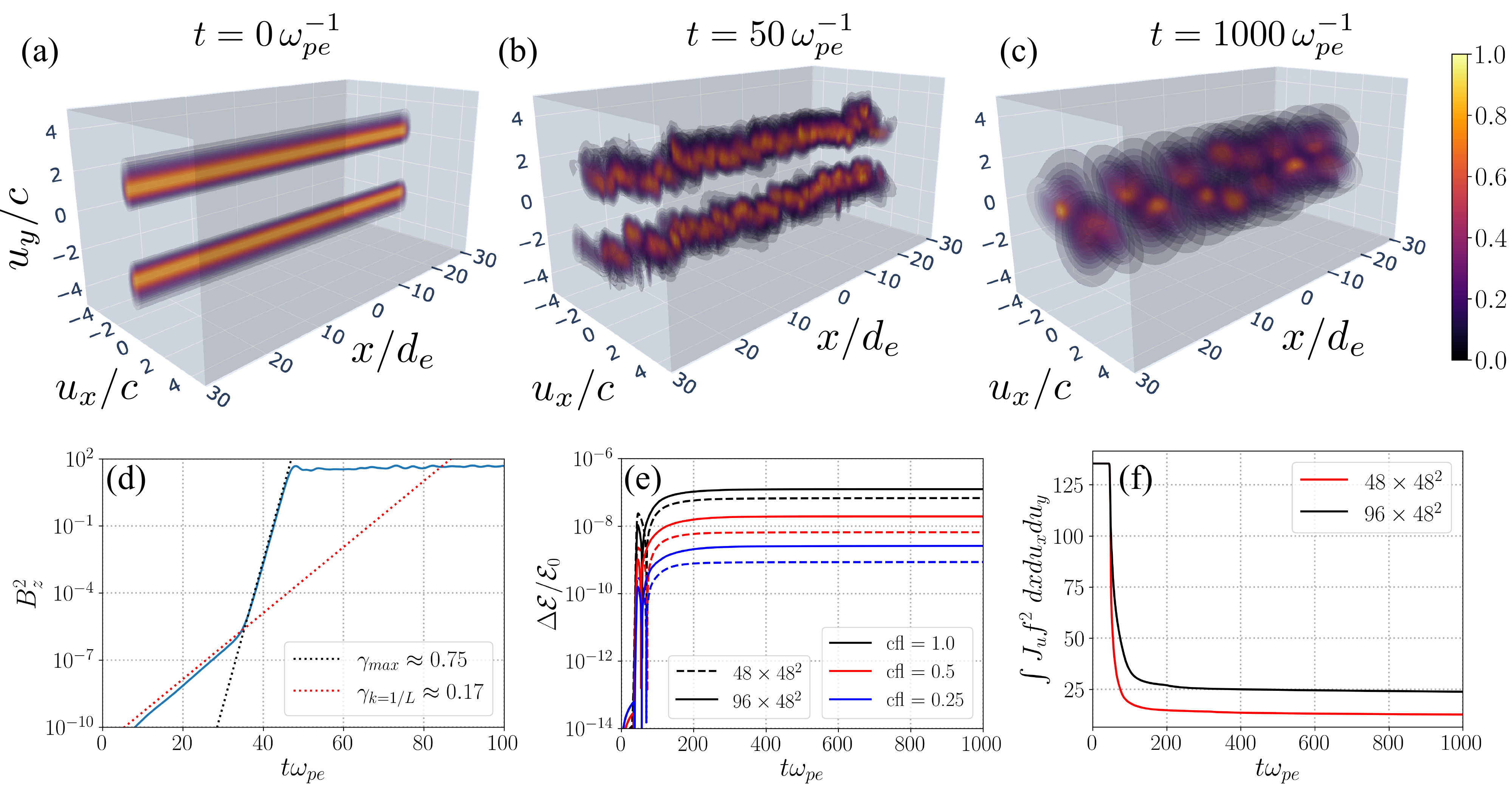}
\caption{Same as Fig.~\ref{fig:TS_evolution}, but for the filamentation instability in the full three-dimensional phase space 
(panel (d) shows the evolution of the energy in the $B_z$ component). 
}\label{fig:Weibel_evolution}
\end{figure*}

We also evolve a pair of simulations deep into the nonlinear regime: a two-stream instability with $k_x d_e = 0.02$ and a filamentation instability with $k_x d_e = 0.1$, to $t = 10^4 \omega_{pe}^{-1}$ and $t = 10^3 \omega_{pe}^{-1}$, respectively. The results are shown in Figures~\ref{fig:TS_evolution} and~\ref{fig:Weibel_evolution}. For each case, we plot the evolution of the distribution function alongside the temporal evolution of the total energy and the $L^2$ energy, $J_u f^2$. The structure of the distribution function in $(x,u_x)$ for the two-stream instability and $(x,u_x,u_y)$ for the filamentation instability agrees with expectations from textbook theory and numerous prior PIC validation studies.

More importantly, we observe the expected behavior of both the total energy and the $L^2$ energy.
Energy conservation depends only on the time-step size and is independent of grid resolution, with errors decreasing as $(\Delta t)^3$, consistent with our use of a third-order, three-stage strong-stability-preserving Runge--Kutta scheme \citep{Shu:2002}. 
The $L^2$ energy decays monotonically throughout the simulations, demonstrating that the scheme remains $L^2$ stable even in the presence of a nonuniform velocity-space discretization.

\subsection{Simulation setup details for electric field screening} \label{sec:discharge}

The \gke\ and \texttt{TRISTAN-MP v2} simulations both extend over a domain $[0, L_x]$ with $L_x = 20$. 
Normalized units are employed, with $\epsilon_0 = \mu_0 = 1$, so that all velocities are expressed in units of the speed of light, $c = 1$. 
Further normalizations are applied for particle properties: $m_e = m_p = 1$, $q_e = -1$, $q_p = 1$, $T_{e,p} = 1$ for electrons and positrons, and a reference density $n_0 = 1$ for both species. 
Consequently, the derived plasma frequency, $\omega_{pe} = \sqrt{q_e^2 n_0/(\epsilon_0 m_e)} = 1$, defines the normalized time scale, while the inertial length, $d_e = c/\omega_{pe} = 1$, sets the normalized length scale.

\gke\ simulations were performed using different configurations of the nonuniform velocity-space map, $N_u = 400, u_{\min} = 0.14$, $N_u = 400, u_{\min} = 0.07$, and $N_u = 800, u_{\min} = 0.14$, with all simulations setting $u_{\max} = 10{,}000$. 
Additionally, \gke\ employs zero-flux boundary conditions in velocity space and a tensor-product polynomial basis with quadratic polynomials in all dimensions for its DG basis. 
For \texttt{TRISTAN-MP v2}, we scanned the initial particles-per-cell from $N_\mathrm{ppc} = 10$ to $N_\mathrm{ppc} = 100$ and varied the time step by scanning the Courant-Friedrichs-Lewy, CFL, number, $\rm{CFL} = 0.45, 0.1$. 
Both \gke\ and \texttt{TRISTAN-MP v2} use $N_x = 200$ and periodic boundary conditions in $x$. 
As stated in the main text, simulations run until $t_\mathrm{end} = 1800 \, \omega_{pe}^{-1}$ with a pair injection rate of $\dot{n} = 0.5 \, n_0 \, \omega_{pe}$. 
The temperature of the injected pairs is set equal to that of the initial pair plasma, $T_\mathrm{source} = 1$. 

The distribution function for both initial and injected pairs follows the Maxwell-Jüttner distribution, with \gke's DG construction following \cite{Johnson:2025} and \texttt{TRISTAN-MP v2}'s sampling procedure following \cite{Zenitani:2015}. Initial perturbations are seeded in the density as
\begin{align}
    n_{e,p}(x,t=0) = n_0 \left[ 1 \pm \sum_{j=1}^5 \xi_j \cos \left( \frac{2\pi j}{L_x} x + \phi_j \right) \right],
\end{align}
where the $\pm$ corresponds to electron and positron perturbations, respectively, $\xi_j = \mathcal{U}(0,\alpha)$ and $\phi_j = \mathcal{U}(0, 2\pi)$ are uniformly sampled random amplitudes and phases, with $\alpha = 0.2$ and $j = 1,\dots,5$.  

The electric field is initialized to satisfy $\nabla_{\mathbf{x}} \cdot \mathbf{E} = \rho_c / \epsilon_0$, with superimposed perturbations on top of a uniform background:
\begin{align}
  {E}(x) = E_0 - 2 \sum_{j=1}^5 \xi_j \frac{L_x}{2\pi j} \sin \left( \frac{2\pi j}{L_x} x + \phi_j \right),
\end{align}
where $E_0 = 400$.

\subsection{Simulation setup details for magnetic reconnection} \label{sec:recon}
The geometry of a force-free current sheet in two spatial dimensions is given by
\begin{align}
    J_{x_{e,p}} & = -\frac{1}{2}\frac{B_0}{w_0} \frac{\tanh \left (\frac{y}{w_0} \right ) \sech^2 \left (\frac{y}{w_0} \right )}{\sqrt{ \left (\frac{B_g}{B_0} \right )^2 + \sech^2 \left (\frac{y}{w_0} \right )}}, \\
    J_{z_{e,p}} & = \frac{1}{2}\frac{B_0}{w_0} \sech^2 \left (\frac{y}{w_0} \right ), \\ 
    B_x & = -B_0 \tanh \left (\frac{y}{w_0} \right ), \\
    B_z & = B_0 \sqrt{\left (\frac{B_g}{B_0} \right )^2 + \sech^2 \left (\frac{y}{w_0} \right )},
\end{align}
where the factor of $1/2$ arises because the current is evenly split between the electrons and positrons to satisfy Ampere's Law. 
A $\sim 1\%$ GEM-like perturbation is utilized \citep{Birn:2001}, along with perturbations to the first 20 wave modes with $\sim 1\%$ noise in $B_x, B_y,$ and $J_z$ to break the symmetry of the GEM-like perturbation and accelerate the development of the reconnection, similar to \cite{TenBarge:2014}.
The guide field strength is set to $B_g = 0.01 B_0$ and the initial layer width is set to $w_0 = d_e$. 
The domain is extended along the length of the current sheet $L_x \times L_y = 64\pi d_e \times 8\pi d_e$ with $N_x \times N_y = 448 \times 56$ grid cells. 
We use the same normalizations as the previous simulations and again choose a reference density $n_0 = 1$ for the electrons and positrons and $T = 0.1$ for the temperatures of both species. 
The value of $B_0$ for the in-plane field strength is determined from the magnetization of plasma, $\sigma = B_0^2/\mu_0 n_0 m_e c^2 = 1$. 

The distribution functions are initialized as Maxwell-Juttner distributions in three velocity dimensions and we choose $N_u = 16, u_{\rm min} = 1,$ and $u_{\rm max} = 32$ in each of the three velocity dimensions with the linear to quadratice nonuniform velocity-space map. 
A tensor-product quadratic polynomial basis is chosen in all five dimensions. 
Periodic boundary conditions are used in $x$, an in-flow boundary condition is used in $y$ to continually fuel the plasma, and zero-flux boundary conditions are used in all three velocity space dimensions. 
We run the simulation for approximately one light-crossing time, $t_{\rm end} = 200 \omega_{pe}^{-1}$.

\bibliography{abbrev,nature_sr_vlasov}

@string{apj =     "Astrophys.~J."}

@string{apjl =    "Astrophys.~J.~Lett."}

@string{cpc =     "Comp.~Phys.~Comm."}

@string{jcomp =   "J.~Comp.~Phys."}

@string{jgr =     "J.~Geophys.~Res."}

@string{jpp = 	  "J.~Plasma Phys."}

@string{mnras =   "Mon.~Not.~Roy.~Astron.~Soc."}

@string{nf =      "Nuc.~Fus."}

@string{pf =      "Phys.~Fluids"}

@string{pop =     "Phys.~Plasmas"}

@string{ppcf = 	  "Plasma Phys.~Con.~Fus."}

@string{prl =     "Phys.~Rev.~Lett."}

@string{s =       "Sci."}

@article{AllmanRahn:2022,
author = {Allmann-Rahn, F. and Lautenbach, S. and Grauer, R.},
title = {An Energy Conserving Vlasov Solver That Tolerates Coarse Velocity Space Resolutions: Simulation of MMS Reconnection Events},
journal = jgr,
volume = {127},
number = {2},
pages = {e2021JA029976},
doi = {https://doi.org/10.1029/2021JA029976},
url = {https://agupubs.onlinelibrary.wiley.com/doi/abs/10.1029/2021JA029976},
year = {2022}
}

@article{Ambrus:2022,
	author = {Ambru{\c s}, V. E. and Bazzanini, L. and Gabbana, A. and Simeoni, D. and Succi, S. and Tripiccione, R.},
	date = {2022/10/01},
	doi = {10.1038/s43588-022-00333-x},
	id = {Ambru{\c s}2022},
	isbn = {2662-8457},
	journal = {Nature Computational Science},
	number = {10},
	pages = {641--654},
	title = {Fast kinetic simulator for relativistic matter},
	url = {https://doi.org/10.1038/s43588-022-00333-x},
	volume = {2},
	year = {2022}
}

@Book{birdsallbook,
  author = 	 {Birdsall, C.K and Langdon, A. B},
  title = 	 {{Plasma Physics Via Computer Simulation}},
  publisher = 	 {Institute of Physics Publishing},
  year = 	 {1990},
}

@article{Birn:2001,
author = {Birn, J. and Drake, J. F. and Shay, M. A. and Rogers, B. N. and Denton, R. E. and Hesse, M. and Kuznetsova, M. and Ma, Z. W. and Bhattacharjee, A. and Otto, A. and Pritchett, P. L.},
title = {Geospace Environmental Modeling (GEM) Magnetic Reconnection Challenge},
journal = jgr,
volume = {106},
number = {A3},
pages = {3715-3719},
doi = {https://doi.org/10.1029/1999JA900449},
url = {https://agupubs.onlinelibrary.wiley.com/doi/abs/10.1029/1999JA900449},
eprint = {https://agupubs.onlinelibrary.wiley.com/doi/pdf/10.1029/1999JA900449},
year = {2001}
}

@article{Bohm:1949,
  title = {Theory of Plasma Oscillations. A. Origin of Medium-Like Behavior},
  author = {Bohm, D. and Gross, E. P.},
  journal = {Phys. Rev.},
  volume = {75},
  issue = {12},
  pages = {1851--1864},
  numpages = {0},
  year = {1949},
  month = {Jun},
  publisher = {American Physical Society},
  doi = {10.1103/PhysRev.75.1851},
  url = {https://link.aps.org/doi/10.1103/PhysRev.75.1851}
}

@article{Bowers:2009,
	doi = {10.1088/1742-6596/180/1/012055},
	url = {https://doi.org/10.1088%2F1742-6596%2F180%2F1%2F012055},
	year = 2009,
	month = {jul},
	publisher = {{IOP} Publishing},
	volume = {180},
	pages = {012055},
	author = {K J Bowers and B J Albright and L Yin and W Daughton and V Roytershteyn and B Bergen and T J T Kwan},
	title = {Advances in petascale kinetic plasma simulation with {VPIC} and Roadrunner},
	journal = {Journal of Physics: Conference Series},
}

@article{Bransgrove:2023,
doi = {10.3847/2041-8213/ad0556},
url = {https://doi.org/10.3847/2041-8213/ad0556},
year = {2023},
month = {nov},
publisher = {The American Astronomical Society},
volume = {958},
number = {1},
pages = {L9},
author = {Bransgrove, Ashley and Beloborodov, Andrei M. and Levin, Yuri},
title = {Radio Emission and Electric Gaps in Pulsar Magnetospheres},
journal = apjl,
}

@article{Breizman:2019,
doi = {10.1088/1741-4326/ab1822},
url = {https://doi.org/10.1088/1741-4326/ab1822},
year = {2019},
month = {jun},
publisher = {IOP Publishing},
volume = {59},
number = {8},
pages = {083001},
author = {Breizman, Boris N. and Aleynikov, Pavel and Hollmann, Eric M. and Lehnen, Michael},
title = {Physics of runaway electrons in tokamaks},
journal = nf,
}

@article{Bret:2009,
	doi = {10.1088/0004-637x/699/2/990},
	url = {https://doi.org/10.1088%2F0004-637x%2F699%2F2%2F990},
	year = {2009},
	month = {jun},
	publisher = {{IOP} Publishing},
	volume = {699},
	number = {2},
	pages = {990--1003},
	author = {A. Bret},
	title = {Weibel, Two--Stream, Filamentation, Oblique, Bell, Buneman... Which One Grows Faster?},
	journal = apj,
}

@article{Bret:2010,
    author = {Bret, A. and Gremillet, L. and Dieckmann, M. E.},
    title = {Multidimensional electron beam-plasma instabilities in the relativistic regime},
    journal = pop,
    volume = {17},
    number = {12},
    pages = {120501},
    year = {2010},
    month = {12},
    issn = {1070-664X},
    doi = {10.1063/1.3514586},
    url = {https://doi.org/10.1063/1.3514586},
}

@article{Buneman:1958,
  title = {Instability, Turbulence, and Conductivity in Current-Carrying Plasma},
  author = {Buneman, O.},
  journal = prl,
  volume = {1},
  issue = {1},
  pages = {8--9},
  numpages = {0},
  year = {1958},
  month = {Jul},
  publisher = {American Physical Society},
  doi = {10.1103/PhysRevLett.1.8},
  url = {https://link.aps.org/doi/10.1103/PhysRevLett.1.8}
}

@article{Cerutti:2013,
doi = {10.1088/0004-637X/770/2/147},
url = {https://doi.org/10.1088/0004-637X/770/2/147},
year = {2013},
month = {jun},
publisher = {The American Astronomical Society},
volume = {770},
number = {2},
pages = {147},
author = {Cerutti, B. and Werner, G. R. and Uzdensky, D. A. and Begelman, M. C.},
title = {SIMULATIONS OF PARTICLE ACCELERATION BEYOND THE CLASSICAL SYNCHROTRON BURNOFF LIMIT IN MAGNETIC RECONNECTION: AN EXPLANATION OF THE CRAB FLARES},
journal = apj,
}

@article{Cheng:2014b,
title = {{Discontinuous Galerkin Methods for the Vlasov-Maxwell Equations}},
journal = {SIAM Journal on Numerical Analysis},
volume = {52},
number = {2},
pages = {1017--1049},
year = {2014},
note = {},
issn = {},
doi = {10.1137/130915091},
author = {Cheng, Yingda and Gamba, Irene M. and Li, Fengyan and Morrison, Philip J.},
}

@article{Chernoglazov:2024,
doi = {10.3847/2041-8213/ad7e24},
url = {https://doi.org/10.3847/2041-8213/ad7e24},
year = {2024},
month = {oct},
publisher = {The American Astronomical Society},
volume = {974},
number = {2},
pages = {L32},
author = {Chernoglazov, Alexander and Philippov, Alexander and Timokhin, Andrey},
title = {Coherence of Multidimensional Pair Production Discharges in Polar Caps of Pulsars},
journal = apjl,
}

@article{Conley:2024,
    author = {Conley, S. A. and Juno, J. and TenBarge, J. M. and Barbhuiya, M. H. and Cassak, P. A. and Howes, G. G. and Lichko, E.},
    title = {The kinetic analog of the pressure--strain interaction},
    journal = pop,
    volume = {31},
    number = {12},
    pages = {122117},
    year = {2024},
    month = {12},
    issn = {1070-664X},
    doi = {10.1063/5.0231200},
    url = {https://doi.org/10.1063/5.0231200},
}

@article{Delzanno:2015,
title = {Multi-dimensional, fully-implicit, spectral method for the Vlasov--Maxwell equations with exact conservation laws in discrete form},
journal = jcomp,
volume = {301},
pages = {338-356},
year = {2015},
issn = {0021-9991},
doi = {https://doi.org/10.1016/j.jcp.2015.07.028},
url = {https://www.sciencedirect.com/science/article/pii/S0021999115004738},
author = {G.L. Delzanno},
}

@InProceedings{Fonseca:2002,
author="Fonseca, R. A.
and Silva, L. O.
and Tsung, F. S.
and Decyk, V. K.
and Lu, W.
and Ren, C.
and Mori, W. B.
and Deng, S.
and Lee, S.
and Katsouleas, T.
and Adam, J. C.",
editor="Sloot, Peter M. A.
and Hoekstra, Alfons G.
and Tan, C. J. Kenneth
and Dongarra, Jack J.",
title="OSIRIS: A Three-Dimensional, Fully Relativistic Particle in Cell Code for Modeling Plasma Based Accelerators",
booktitle="Computational Science --- ICCS 2002",
year="2002",
publisher="Springer Berlin Heidelberg",
address="Berlin, Heidelberg",
pages="342--351",
isbn="978-3-540-47789-1"
}

@article{Fonseca:2008,
	doi = {10.1088/0741-3335/50/12/124034},
	url = {https://doi.org/10.1088%2F0741-3335%2F50%2F12%2F124034},
	year = 2008,
	month = {nov},
	publisher = {{IOP} Publishing},
	volume = {50},
	number = {12},
	pages = {124034},
	author = {R A Fonseca and S F Martins and L O Silva and J W Tonge and F S Tsung and W B Mori},
	title = {One-to-one direct modeling of experiments and astrophysical scenarios: pushing the envelope on kinetic plasma simulations},
	journal = ppcf,
}

@misc{Francisquez:2026,
      title={Conservative velocity mappings for discontinuous Galerkin kinetics}, 
      author={Manaure Francisquez and Petr Cagas and Akash Shukla and James Juno and Gregory W. Hammett},
      year={2026},
      eprint={2505.10754},
      archivePrefix={arXiv},
      primaryClass={physics.plasm-ph},
      url={https://arxiv.org/abs/2505.10754}, 
}

@article{Fried:1959,
author = {Fried, Burton D. },
title = {Mechanism for Instability of Transverse Plasma Waves},
journal = pf,
volume = {2},
number = {3},
pages = {337-337},
year = {1959},
doi = {10.1063/1.1705933},
URL = {https://aip.scitation.org/doi/abs/10.1063/1.1705933},
}

@misc{Galishnikova:2025,
      title={$\mathtt{Entity}$ -- Hardware-agnostic Particle-in-Cell Code for Plasma Astrophysics. II: General Relativistic Module}, 
      author={Alisa Galishnikova and Hayk Hakobyan and Alexander Philippov and Benjamin Crinquand},
      year={2025},
      eprint={2511.17701},
      archivePrefix={arXiv},
      primaryClass={astro-ph.HE},
      url={https://arxiv.org/abs/2511.17701}, 
}

@article{Germaschewski:2016,
title = {The Plasma Simulation Code: A modern particle-in-cell code with patch-based load-balancing},
journal = jcomp,
volume = {318},
pages = {305--326},
year = {2016},
issn = {0021-9991},
doi = {https://doi.org/10.1016/j.jcp.2016.05.013},
url = {http://www.sciencedirect.com/science/article/pii/S0021999116301413},
author = {Kai Germaschewski and William Fox and Stephen Abbott and Narges Ahmadi and Kristofor Maynard and Liang Wang and Hartmut Ruhl and Amitava Bhattacharjee},
}

@article{Gonoskov:2022,
  title = {Charged particle motion and radiation in strong electromagnetic fields},
  author = {Gonoskov, A. and Blackburn, T. G. and Marklund, M. and Bulanov, S. S.},
  journal = {Rev. Mod. Phys.},
  volume = {94},
  issue = {4},
  pages = {045001},
  numpages = {63},
  year = {2022},
  month = {Oct},
  publisher = {American Physical Society},
  doi = {10.1103/RevModPhys.94.045001},
  url = {https://link.aps.org/doi/10.1103/RevModPhys.94.045001}
}

@misc{grunwald:2025,
      title={Solving the six-dimensional Vlasov-Maxwell System with Active Flux and Splitting Methods}, 
      author={G. Gr\"unwald and L. Hensel and M. Deisenhofer and S. Lautenbach and K. Kormann and R. Grauer},
      year={2025},
      eprint={2511.22440},
      archivePrefix={arXiv},
      primaryClass={physics.plasm-ph},
      url={https://arxiv.org/abs/2511.22440}, 
}

@article{Guo:2014,
  title = {Formation of Hard Power Laws in the Energetic Particle Spectra Resulting from Relativistic Magnetic Reconnection},
  author = {Guo, Fan and Li, Hui and Daughton, William and Liu, Yi-Hsin},
  journal = prl,
  volume = {113},
  issue = {15},
  pages = {155005},
  numpages = {5},
  year = {2014},
  month = {Oct},
  publisher = {American Physical Society},
  doi = {10.1103/PhysRevLett.113.155005},
  url = {https://link.aps.org/doi/10.1103/PhysRevLett.113.155005}
}

@misc{Hakim:2019,
      title={Discontinuous Galerkin schemes for a class of Hamiltonian evolution equations with applications to plasma fluid and kinetic problems}, 
      author={A. Hakim and G. Hammett and E. Shi and N. Mandell},
      year={2019},
      eprint={1908.01814},
      archivePrefix={arXiv},
      primaryClass={physics.comp-ph},
      url={https://arxiv.org/abs/1908.01814}, 
}

@inproceedings{HakimJuno:2020,
  author    = {Hakim, Ammar and Juno, James},
  title     = {Alias-Free, Matrix-Free, and Quadrature-Free Discontinuous Galerkin Algorithms for (Plasma) Kinetic Equations},
  booktitle = {Proceedings of the International Conference for High Performance Computing, Networking, Storage and Analysis},
  year      = {2020},
  pages     = {1--15},
  number    = {73},
  publisher = {IEEE Press},
  isbn      = {9781728199986}
}

@INPROCEEDINGS{Hakobyan:2024,
       author = {{Hakobyan}, Hayk and {Spitkovsky}, Anatoly and {Chernoglazov}, Alexander and {Philippov}, Alexander and {Groselj}, Daniel and {Mahlmann}, Jens},
        title = "{PrincetonUniversity/tristan-mp-v2: v2.6}",
    booktitle = {Zenodo},
         year = 2024,
       volume = {75},
        month = jan,
    publisher = {Zenodo},
          eid = {7566725},
        pages = {7566725},
          doi = {10.5281/zenodo.7566725},
       adsurl = {https://ui.adsabs.harvard.edu/abs/2024zndo...7566725H}
}

@misc{Hakobyan:2025,
      title={Entity -- Hardware-agnostic Particle-in-Cell Code for Plasma Astrophysics. I: Curvilinear Special Relativistic Module}, 
      author={Hayk Hakobyan and Ludwig M. B{\"o}ss and Yangyang Cai and Alexander Chernoglazov and Alisa Galishnikova and Evgeny A. Gorbunov and Jens F. Mahlmann and Alexander Philippov and Siddhant Solanki and Arno Vanthieghem and Muni Zhou},
      year={2025},
      eprint={2511.17710},
      archivePrefix={arXiv},
      primaryClass={astro-ph.HE},
      url={https://arxiv.org/abs/2511.17710}, 
}

@article{Harris:2020,
	author = {Harris, Charles R. and Millman, K. Jarrod and van der Walt, St{\'e}fan J. and Gommers, Ralf and Virtanen, Pauli and Cournapeau, David and Wieser, Eric and Taylor, Julian and Berg, Sebastian and Smith, Nathaniel J. and Kern, Robert and Picus, Matti and Hoyer, Stephan and van Kerkwijk, Marten H. and Brett, Matthew and Haldane, Allan and del R{\'\i}o, Jaime Fern{\'a}ndez and Wiebe, Mark and Peterson, Pearu and G{\'e}rard-Marchant, Pierre and Sheppard, Kevin and Reddy, Tyler and Weckesser, Warren and Abbasi, Hameer and Gohlke, Christoph and Oliphant, Travis E.},
	date = {2020/09/01},
	doi = {10.1038/s41586-020-2649-2},
	id = {Harris2020},
	isbn = {1476-4687},
	journal = {Nature},
	number = {7825},
	pages = {357--362},
	title = {Array programming with NumPy},
	url = {https://doi.org/10.1038/s41586-020-2649-2},
	volume = {585},
	year = {2020}
    }

@book{Hesthaven:2007,
  title =	 {Nodal discontinuous {G}alerkin methods: algorithms,
                  analysis, and applications},
  author =	 {Hesthaven, J.S. and Warburton, T.},
  year =	 {2007},
  publisher =	 {Springer Science \& Business Media}
}

@article{Horvath:2020,
    author = {Horvath, Sarah A. and Howes, Gregory G. and McCubbin, Andrew J.},
    title = {Electron Landau damping of kinetic Alfv{\'e}n waves in simulated magnetosheath turbulence},
    journal = pop,
    volume = {27},
    number = {10},
    pages = {102901},
    year = {2020},
    month = {10},
    issn = {1070-664X},
    doi = {10.1063/5.0021727},
    url = {https://doi.org/10.1063/5.0021727},
}

@article{Howes:2017, 
title={Diagnosing collisionless energy transfer using field--particle correlations: Vlasov--Poisson plasmas}, 
volume={83}, 
doi={10.1017/S0022377816001197}, 
number={1}, 
journal=jpp, 
publisher={Cambridge University Press}, 
author={Howes, Gregory G. and Klein, Kristopher G. and Li, Tak Chu}, 
year={2017}, 
pages={705830102},
url = {https://www.cambridge.org/core/article/diagnosing-collisionless-energy-transfer-using-fieldparticle-correlations-vlasovpoisson-plasmas/12921EF6C25C03AC133C892D5A2E4B89},
}

@article{Howes:2025,
    author = {Howes, Gregory G. and Felix, Alberto and Brown, Collin R. and Haggerty, Colby C. and Juno, James and TenBarge, Jason M. and Wilson, Lynn B., III and Caprioli, Damiano},
    title = {Velocity-space signatures of shock-drift acceleration at quasi-perpendicular collisionless shocks},
    journal = pop,
    volume = {32},
    number = {6},
    pages = {062904},
    year = {2025},
    month = {06},
    issn = {1070-664X},
    doi = {10.1063/5.0269528},
    url = {https://doi.org/10.1063/5.0269528},
}

@article{Hunter:2007,
  author={Hunter, John D.},
  journal={Computing in Science \& Engineering}, 
  title={Matplotlib: A 2D Graphics Environment}, 
  year={2007},
  volume={9},
  number={3},
  pages={90-95},
  doi={10.1109/MCSE.2007.55}
  }

@article{Johnson:2025, 
title={A moment-conserving discontinuous Galerkin representation of the relativistic Maxwellian distribution}, 
volume={91}, 
DOI={10.1017/S0022377825100718}, 
number={5}, 
journal=jpp, 
author={Johnson, Grant and Hakim, Ammar and Juno, James}, 
year={2025}, 
pages={E130}
}

@misc{Johnson:2026,
      title={A Conservative Discontinuous Galerkin Algorithm for Particle Kinetics on Smooth Manifolds}, 
      author={Grant Johnson and Ammar Hakim and James Juno},
      year={2025},
      eprint={2512.05298},
      archivePrefix={arXiv},
      primaryClass={physics.comp-ph},
      url={https://arxiv.org/abs/2512.05298}, 
}

@article{Juno:2018,
   author = {{Juno}, J. and {Hakim}, A. and {TenBarge}, J. and {Shi}, E. and 
	{Dorland}, W.},
    title = {{Discontinuous Galerkin algorithms for fully kinetic plasmas}},
  journal = jcomp,
     year = 2018,
    month = {Jan},
   volume = 353,
    pages = {110-147},
      doi = {10.1016/j.jcp.2017.10.009},
    url = {http://www.sciencedirect.com/science/article/pii/S0021999117307477},
}

@article{Juno:2021,
       author = {{Juno}, James and {Howes}, Gregory G. and {TenBarge}, Jason M. and {Wilson}, Lynn B. and {Spitkovsky}, Anatoly and {Caprioli}, Damiano and {Klein}, Kristopher G. and {Hakim}, Ammar},
        title = "{A field-particle correlation analysis of a perpendicular magnetized collisionless shock}",
      journal = jpp,
         year = 2021,
        month = {Jun},
       volume = {87},
       number = {3},
          eid = {905870316},
        pages = {905870316},
          doi = {10.1017/S0022377821000623},
 primaryClass = {physics.plasm-ph},
       adsurl = {https://ui.adsabs.harvard.edu/abs/2021JPlPh..87c9016J}
}

@article{Juno:2023,
	adsurl = {https://ui.adsabs.harvard.edu/abs/2023ApJ...944...15J},
	archiveprefix = {arXiv},
	author = {{Juno}, James and {Brown}, Collin R. and {Howes}, Gregory G. and {Haggerty}, Colby C. and {TenBarge}, Jason M. and {Wilson}, Lynn B., III and {Caprioli}, Damiano and {Klein}, Kristopher G.},
	doi = {10.3847/1538-4357/acaf53},
	eid = {15},
	eprint = {2211.15340},
	journal = apj,
	month = feb,
	number = {1},
	pages = {15},
	primaryclass = {physics.plasm-ph},
	title = {{Phase-space Energization of Ions in Oblique Shocks}},
	volume = {944},
	year = 2023
}

@article{Juno:2025, 
title={A parallel-kinetic-perpendicular-moment model for magnetised plasmas}, 
volume={91}, 
DOI={10.1017/S0022377825100706}, 
number={5}, 
journal=jpp, 
author={Juno, James and Hakim, Ammar and TenBarge, Jason M.}, 
year={2025}, 
pages={E129}
}

@article{Kempf:2013,
author = {Kempf, Yann and Pokhotelov, Dimitry and {Von Alfthan}, Sebastian and Vaivads, Andris and Palmroth, Minna and Koskinen, Hannu E J},
doi = {10.1063/1.4835315},
issn = {1070664X},
journal = pop,
number = {11},
pages = {1--9},
primaryClass = {physics.plasm-ph},
title = {{Wave dispersion in the hybrid-Vlasov model: Verification of Vlasiator}},
volume = {20},
URL = {https://doi.org/10.1063/1.4835315},
year = {2013}
}

@article{Klein:2016,
	doi = {10.3847/2041-8205/826/2/l30},
	url = {https://doi.org/10.3847%2F2041-8205%2F826%2F2%2Fl30},
	year = {2016},
	month = {jul},
	publisher = {American Astronomical Society},
	volume = {826},
	number = {2},
	pages = {L30},
	author = {K. G. Klein and G. G. Howes},
        title = {Measuring Collisionless Damping in Heliospheric Plasmas Using Field--Particle Correlations},
	journal = apjl,
}

@article{Klein:2017b,
title={Diagnosing collisionless energy transfer using field--particle correlations: gyrokinetic turbulence}, 
volume={83}, 
doi={10.1017/S0022377817000563}, 
number={4}, 
journal=jpp, 
publisher={Cambridge University Press}, 
author={Klein, Kristopher G. and Howes, Gregory G. and TenBarge, Jason M.}, 
year={2017}, 
pages={535830401},
url = {https://www.cambridge.org/core/article/diagnosing-collisionless-energy-transfer-using-fieldparticle-correlations-gyrokinetic-turbulence/DDDF5DBB5E4A5CF140F8204A69D9A0A7},
}

@article{Klein:2020,
title={Diagnosing collisionless energy transfer using field--particle correlations: Alfv{\'e}n-ion cyclotron turbulence}, 
volume={86}, 
DOI={10.1017/S0022377820000689}, 
number={4}, 
journal=jpp, 
publisher={Cambridge University Press}, 
author={Klein, Kristopher G. and Howes, Gregory G. and TenBarge, Jason M. and Valentini, Francesco}, 
year={2020}, 
pages={905860402}}

@article{Kormann:2025,
title = {A structure-preserving finite element framework for the Vlasov–Maxwell system},
journal = {Computer Methods in Applied Mechanics and Engineering},
volume = {446},
pages = {118290},
year = {2025},
issn = {0045-7825},
doi = {https://doi.org/10.1016/j.cma.2025.118290},
url = {https://www.sciencedirect.com/science/article/pii/S0045782525005626},
author = {Katharina Kormann and Murtazo Nazarov and Junjie Wen},
}

@article{Koshkarov:2021,
title = {The multi-dimensional Hermite-discontinuous Galerkin method for the Vlasov--Maxwell equations},
journal = cpc,
volume = {264},
pages = {107866},
year = {2021},
issn = {0010-4655},
doi = {https://doi.org/10.1016/j.cpc.2021.107866},
url = {https://www.sciencedirect.com/science/article/pii/S0010465521000266},
author = {O. Koshkarov and G. Manzini and G.L. Delzanno and C. Pagliantini and V. Roytershteyn},
}

@article{Pagliantini:2023,
title = {Energy-conserving explicit and implicit time integration methods for the multi-dimensional Hermite-DG discretization of the Vlasov-Maxwell equations},
journal = cpc,
volume = {284},
pages = {108604},
year = {2023},
issn = {0010-4655},
doi = {https://doi.org/10.1016/j.cpc.2022.108604},
url = {https://www.sciencedirect.com/science/article/pii/S001046552200323X},
author = {C. Pagliantini and G. Manzini and O. Koshkarov and G.L. Delzanno and V. Roytershteyn},
}

@article{Palmroth:2018,
Author = {Palmroth, Minna and Ganse, Urs and Pfau-Kempf, Yann and Battarbee, Markus and Turc, Lucile and Brito, Thiago and Grandin, Maxime and Hoilijoki, Sanni and Sandroos, Arto and von Alfthan, Sebastian},
Da = {2018/08/16},
Doi = {10.1007/s41115-018-0003-2},
Isbn = {2365-0524},
Journal = {Living Reviews in Computational Astrophysics},
Number = {1},
Pages = {1},
Title = {Vlasov methods in space physics and astrophysics},
Url = {https://doi.org/10.1007/s41115-018-0003-2},
Volume = {4},
Year = {2018},
}

@article{Parfrey:2019,
  title = {First-Principles Plasma Simulations of Black-Hole Jet Launching},
  author = {Parfrey, Kyle and Philippov, Alexander and Cerutti, Beno\^{\i}t},
  journal = prl,
  volume = {122},
  issue = {3},
  pages = {035101},
  numpages = {6},
  year = {2019},
  month = {Jan},
  publisher = {American Physical Society},
  doi = {10.1103/PhysRevLett.122.035101},
  url = {https://link.aps.org/doi/10.1103/PhysRevLett.122.035101}
}

@article{Philippov:2020,
  title = {Origin of Pulsar Radio Emission},
  author = {Philippov, Alexander and Timokhin, Andrey and Spitkovsky, Anatoly},
  journal = prl,
  volume = {124},
  issue = {24},
  pages = {245101},
  numpages = {5},
  year = {2020},
  month = {Jun},
  publisher = {American Physical Society},
  doi = {10.1103/PhysRevLett.124.245101},
  url = {https://link.aps.org/doi/10.1103/PhysRevLett.124.245101}
}

@article{Philippov:2022,
author = {Philippov, A. and Kramer, M.},
title = {Pulsar Magnetospheres and Their Radiation},
journal = {Annual Review of Astronomy and Astrophysics},
volume = {60},
number = {1},
pages = {495-558},
year = {2022},
doi = {10.1146/annurev-astro-052920-112338},
}

@article{Roytershteyn:2018,
author={Roytershteyn, Vadim and Delzanno, Gian Luca},   	 
title={Spectral Approach to Plasma Kinetic Simulations Based on Hermite Decomposition in the Velocity Space},      	
journal={Frontiers in Astron.~Space~Sci.},	
volume={5},
pages={27},	
year={2018},	  
doi={10.3389/fspas.2018.00027},      
}

@article{Shu:2002,
  title =	 {A survey of strong stability preserving high order
                  time discretizations},
  author =	 {Shu, C.-W.},
  journal =	 {Collected lectures on the preservation of stability
                  under discretization},
  volume =	 {109},
  pages =	 {51--65},
  year =	 {2002},
  publisher =	 {SIAM (Philadelphia, PA)}
}

@article{Sironi:2014,
doi = {10.1088/2041-8205/783/1/L21},
url = {https://doi.org/10.1088/2041-8205/783/1/L21},
year = {2014},
month = {feb},
publisher = {The American Astronomical Society},
volume = {783},
number = {1},
pages = {L21},
author = {Sironi, Lorenzo and Spitkovsky, Anatoly},
title = {RELATIVISTIC RECONNECTION: AN EFFICIENT SOURCE OF NON-THERMAL PARTICLES},
journal = apjl,
}

@inproceedings{Spitkovsky:2005,
       author = {{Spitkovsky}, Anatoly},
        title = {{Simulations of relativistic collisionless shocks: shock structure and particle acceleration}},
    booktitle = {Astrophysical Sources of High Energy Particles and Radiation},
         year = 2005,
       editor = {{Bulik}, Tomasz and {Rudak}, Bronislaw and {Madejski}, Grzegorz},
       series = {American Institute of Physics Conference Series},
       volume = {801},
        month = nov,
        pages = {345--350},
          doi = {10.1063/1.2141897},
       adsurl = {https://ui.adsabs.harvard.edu/abs/2005AIPC..801..345S}
}

@article{TenBarge:2014,
    author = {TenBarge, J. M. and Daughton, W. and Karimabadi, H. and Howes, G. G. and Dorland, W.},
    title = {Collisionless reconnection in the large guide field regime: Gyrokinetic versus particle-in-cell simulations},
    journal = pop,
    volume = {21},
    number = {2},
    pages = {020708},
    year = {2014},
    month = {02},
    issn = {1070-664X},
    doi = {10.1063/1.4867068},
    url = {https://doi.org/10.1063/1.4867068},
    eprint = {https://pubs.aip.org/aip/pop/article-pdf/doi/10.1063/1.4867068/15742188/020708\_1\_online.pdf},
}

@article{Timokhin:2010,
    author = {Timokhin, A. N.},
    title = {Time-dependent pair cascades in magnetospheres of neutron stars -- I. Dynamics of the polar cap cascade with no particle supply from the neutron star surface},
    journal = mnras,
    volume = {408},
    number = {4},
    pages = {2092-2114},
    year = {2010},
    month = {10},
    issn = {0035-8711},
    doi = {10.1111/j.1365-2966.2010.17286.x},
    url = {https://doi.org/10.1111/j.1365-2966.2010.17286.x},
    eprint = {https://academic.oup.com/mnras/article-pdf/408/4/2092/4220523/mnras0408-2092.pdf},
}

@article{Timokhin:2013,
    author = {Timokhin, A. N. and Arons, J.},
    title = {Current flow and pair creation at low altitude in rotation-powered pulsars' force-free magnetospheres: space charge limited flow},
    journal = mnras,
    volume = {429},
    number = {1},
    pages = {20-54},
    year = {2012},
    month = {12},
    issn = {0035-8711},
    doi = {10.1093/mnras/sts298},
    url = {https://doi.org/10.1093/mnras/sts298},
    eprint = {https://academic.oup.com/mnras/article-pdf/429/1/20/18684664/sts298.pdf},
}

@article{Tolman:2022,
doi = {10.3847/2041-8213/ac7c71},
url = {https://dx.doi.org/10.3847/2041-8213/ac7c71},
year = {2022},
month = {Jul},
publisher = {The American Astronomical Society},
volume = {933},
number = {2},
pages = {L37},
author = {Elizabeth A. Tolman and A. A. Philippov and A. N. Timokhin},
title = {Electric Field Screening in Pair Discharges and Generation of Pulsar Radio Emission},
journal = apj,
}

@article{Valentini:2007,
Author = {{Valentini}, F. and {Tr{\'a}vn{\'{\i}}{\v c}ek}, P. and {Califano}, F. and {Hellinger}, P. and {Mangeney}, A.},
Doi = {10.1016/j.jcp.2007.01.001},
Journal = jcomp,
Month = jul,
Pages = {753-770},
Title = {{A hybrid-Vlasov model based on the current advance method for the simulation of collisionless magnetized plasma}},
Volume = 225,
Year = 2007,
url = {http://www.sciencedirect.com/science/article/pii/S0021999107000022},
}

@article{Vencels:2016,
doi = {10.1088/1742-6596/719/1/012022},
url = {https://doi.org/10.1088%2F1742-6596%2F719%2F1%2F012022},
year = 2016,
month = {may},
publisher = {{IOP} Publishing},
volume = {719},
pages = {012022},
author = {Juris Vencels and Gian Luca Delzanno and Gianmarco Manzini and Stefano Markidis and Ivy Bo Peng and Vadim Roytershteyn},
title = {{SpectralPlasmaSolver}: a Spectral Code for Multiscale Simulations of Collisionless, Magnetized Plasmas},
journal = {J.~Phys.~Conference Series},
}

@article{Weibel:1959,
  title = {Spontaneously Growing Transverse Waves in a Plasma Due to an Anisotropic Velocity Distribution},
  author = {Weibel, Erich S.},
  journal = prl,
  volume = {2},
  issue = {3},
  pages = {83--84},
  numpages = {0},
  year = {1959},
  month = {Feb},
  doi = {10.1103/PhysRevLett.2.83},
  url = {https://link.aps.org/doi/10.1103/PhysRevLett.2.83}
}

@article{Werner:2017,
doi = {10.3847/2041-8213/aa7892},
url = {https://doi.org/10.3847/2041-8213/aa7892},
year = {2017},
month = {jul},
publisher = {The American Astronomical Society},
volume = {843},
number = {2},
pages = {L27},
author = {Werner, Gregory R. and Uzdensky, Dmitri A.},
title = {Nonthermal Particle Acceleration in 3D Relativistic Magnetic Reconnection in Pair Plasma},
journal = apjl,
}

@article{Wettervik:2017,
Author = {Wettervik, Benjamin Svedung and DuBois, Timothy C. and Siminos, Evangelos and F{\"u}l{\"o}p, T{\"u}nde},
Da = {2017/06/15},
Doi = {10.1140/epjd/e2017-80102-2},
Isbn = {1434-6079},
Journal = {The European Phys.~J.~D},
Number = {6},
Pages = {157},
Title = {Relativistic Vlasov-Maxwell modelling using finite volumes and adaptive mesh refinement},
Url = {https://doi.org/10.1140/epjd/e2017-80102-2},
Volume = {71},
Year = {2017},
}

@article{Ye:2025,
doi = {10.3847/1538-4357/ae19e5},
url = {https://doi.org/10.3847/1538-4357/ae19e5},
year = {2025},
month = {dec},
publisher = {The American Astronomical Society},
volume = {995},
number = {2},
pages = {179},
author = {Ye, Dingyi and Chen, Alexander Y.},
title = {1D Vlasov Simulations of QED Cascades over Pulsar Polar Caps},
journal = apj,
}

@article{Zeng:2026,
doi = {10.3847/2041-8213/ae2ade},
url = {https://doi.org/10.3847/2041-8213/ae2ade},
year = {2025},
month = {dec},
publisher = {The American Astronomical Society},
volume = {996},
number = {2},
pages = {L20},
author = {Zeng, Shuzhe and Philippov, Alexander and Juno, James and Beloborodov, Andrei M. and Popova, Elena},
title = {Origin of Pulsed Radio Emission from Magnetars},
journal = apjl,
}

@article{Zenitani:2015,
    author = {Zenitani, Seiji},
    title = {Loading relativistic Maxwell distributions in particle simulations},
    journal = pop,
    volume = {22},
    number = {4},
    pages = {042116},
    year = {2015},
    month = {04},
    issn = {1070-664X},
    doi = {10.1063/1.4919383},
    url = {https://doi.org/10.1063/1.4919383},
}

\end{document}